\documentclass[useAMS,usenatbib]{mn2e}
\usepackage{natbib}
\usepackage{times}
\usepackage{rotate}
\usepackage{color}

\usepackage{graphicx}
\usepackage{amsmath,amssymb}

\newcommand{\THE}{{\boldsymbol{\theta}}}

\newcommand{\ELL}{{\boldsymbol{\ell}}}

\newcommand{\LAMBDA}{{\boldsymbol{\lambda}}}
\newcommand{\bq}{\boldsymbol{q}}

\newcommand{\ave}[1]{\left\langle #1\right\rangle}

\newcommand{\BF}{\begin{figure}\begin{center}}
\newcommand{\EF}{\end{center}\end{figure}}
\newcommand{\BE}{\begin{equation}}
\newcommand{\EE}{\end{equation}}
\newcommand{\BEA}{\begin{eqnarray}}
\newcommand{\EEA}{\end{eqnarray}}

\newcommand{\simgt}{\lower.5ex\hbox{$\; \buildrel > \over \sim \;$}}
\newcommand{\simlt}{\lower.5ex\hbox{$\; \buildrel < \over \sim \;$}}
\newcommand{\rmd}{\ensuremath{\mathrm{d}}}
\newcommand{\bm}[1]{{\boldsymbol{#1}}}

\begin{document}
\title[An optimal survey geometry of WL survey]{An optimal survey geometry of weak lensing survey: minimizing
super-sample covariance}
\author[R. Takahashi, S. Soma, et al.]
{Ryuichi Takahashi$^1$, Shunji Soma$^1$, Masahiro Takada$^2$ and Issha Kayo$^{3,4}$
\\
$^{1}$Faculty of Science and Technology, Hirosaki University,
 3 Bunkyo-cho, Hirosaki, Aomori 036-8561, Japan
\\
$^{2}$Kavli Institute for the Physics and Mathematics of the Universe
 (Kavli IPMU, WPI), The University of Tokyo, Chiba 277-8582, Japan
\\
$^{3}$Department of Physics, Toho University, 2-2-1 Miyama, Funabashi, 
Chiba 274-8510, Japan
\\
$^{4}$Department of Liberal Arts, Tokyo University of Technology, 5-23-22
Nishikamata, Ota-ku, Tokyo 114-8650, Japan
}

\date{\today}


\maketitle

\label{firstpage}
\begin{abstract}
Upcoming wide-area weak lensing surveys are expensive both in time and
cost and require an optimal survey design in order to attain maximum
scientific returns from a fixed amount of available telescope time. The
super-sample covariance (SSC), which arises from unobservable modes that
are larger than the survey size, significantly degrades the statistical 
precision of weak 
lensing power spectrum measurement even for a wide-area survey. 
Using the 1000 mock realizations of
the log-normal model, which approximates the weak lensing field for a
$\Lambda$-dominated cold dark matter model, we study an optimal survey
geometry to minimize the impact of SSC contamination. 
For a continuous survey geometry with a fixed survey area,
a more elongated geometry such as a rectangular shape of 1:400
side-length ratio reduces the SSC effect and allows for
a factor 2 improvement in the cumulative signal-to-noise ratio ($S/N$)
of power spectrum measurement up to $\ell_{\rm max}\simeq $ a few $10^3$,
compared to compact geometries such as squares or circles.
When we allow the survey geometry to be disconnected but with a fixed
 total area, assuming $1\times 1$ sq. degrees patches as the fundamental
 building blocks of survey footprints, 
the best strategy is to locate the patches with $\sim 15$ degrees separation.
This separation angle corresponds to the scale at which 
the two-point correlation function has a negative minimum. 
The best configuration allows for
a factor 100 gain in the effective area coverage as well as a factor 2.5
improvement in the $S/N$ at high multipoles, yielding a much wider
coverage of multipoles than in the compact geometry.
\end{abstract}

\begin{keywords}
cosmology: theory - gravitational lensing: weak -
 large-scale structure of the universe 
\end{keywords}

\section{Introduction}
Weak gravitational
 lensing of foreground large-scale structure induces a coherent,
 correlated distortion 
in distant galaxy images, 
the so-called cosmic shear
~\citep[e.g.,][]{bs01,HoekstraJain:08,munshi08}.
The cosmic shear signal is statistically measurable, e.g., by
measuring the angular two-point correlation function of galaxy images. 
The current state-of-art measurements are from
the Canada-France Hawaii Telescope Lensing
Survey \citep[CFHTLenS;][]{kilb13,heym13} and the Sloan Digital Sky
Survey \citep[SDSS;][]{Huffetal:14,lin12,Mandelbaumetal:13}, which were
used to constrain cosmological parameters such as 
 the present-day amplitude of
 density fluctuation $\sigma_8$ and the matter density parameter 
 $\Omega_{\rm m}$.
There are various on-going and planned surveys aimed at achieving the
high precision measurement such as
the Subaru Hyper Suprime-Cam \citep[HSC;][]{miyazaki06}\footnote{http://www.naoj.org/Projects/HSC/index.html},
the Kilo Degree Survey (KiDS)\footnote{http://kids.strw.leidenuniv.nl},
 the Dark Energy Survey (DES)\footnote{http://www.darkenergysurvey.org/},
 the Panoramic Survey Telescope and Rapid Response
 System (Pan-STARRS)\footnote{http://ps1sc.org/}, and then ultimately
 the Large Synoptic Survey Telescope
 (LSST)\footnote{http://www.lsst.org/lsst/},
the Euclid \footnote{http://sci.esa.int/euclid/}
and the WFIRST \citep{WFIRST}.

Upcoming wide-area galaxy surveys are expensive both in time and cost.
To attain the full potential of the galaxy surveys for a limited amount of available telescope time, it is important to
explore an optimal survey design.
The
statistical precision of the cosmic shear two-point correlation function
or the Fourier-transformed counterpart, the power spectrum, is
determined by their covariance matrix that itself contains two
contributions; the shape noise and the sample variance caused by an
incomplete sampling of the fluctuations due to a finite-area survey.

Even though the initial density field is nearly Gaussian, the sample
variance of large-scale structure probes (here cosmic shear) gets
substantial non-Gaussian contributions from the nonlinear evolution of
large-scale structure
\citep{MeiksinWhite:99,szh99,HuWhite:01,ch01}. Most of the useful
information in the cosmic shear signal lies in the deeply nonlinear
regime \citep{JainSeljak:97,Bernardeauetal:97}.  The super-sample
covariance (SSC) is the sampling variance caused by coupling of
short-wavelength modes relevant for the power spectrum measurement with
very long-wavelength modes larger than the survey size \citep{th13}. It
has been shown to be the largest non-Gaussian contribution to the power
spectrum covariance over a wide range of modes from the weakly to deeply
nonlinear regime
\citep{nsr06,ns07,lp08,TakadaJain:09,taka09,sato09,taka11a,dep12,Kayoetal:13,TakadaSpergel:13,Lietal:14,Lietal:14b}
\citep[see also][for the pioneering work]{HuKravtsov:03,rh05,hrs06}.  The
SSC depends on a survey geometry through the variance of average
convergence mode within the survey window, $(\sigma_W)^2\equiv
\ave{\bar{\kappa}_W^2}$ (here $\bar{\kappa}_W$ is the mean convergence
averaged within the survey area), which differs from the dependence of
other covariance terms that scale as $1/\Omega_S$ ($\Omega_S$ is the
survey area).

The purpose of this paper is to study an optimal survey strategy
for the lensing power spectrum measurement taking account of the SSC
contamination. We study both cases of continuous geometry and sparse
sampling strategy. 
Although previous works showed that the sparse sampling helps
to reduce the sample variance and to give an access to
larger-angle scales than in a continuous geometry \citep{kaiser86,kaiser98,ks04}
\citep[also see][for the similar discussion on the galaxy clustering
analysis]{blake06,pj13,ch13}, we here pay particular attention to
optimization of survey geometry to minimize the SSC contamination. 
To study these issues, we use realizations of weak lensing convergence
field constructed based on the log-normal model, which approximately
describes the weak lensing field for a $\Lambda$CDM model \citep[also
see][for the similar study along the galaxy redshift survey]{ch13}.
For the log-normal model we can analytically derive the power spectrum
covariance following the formulation in \citet{th13}, and use the
analytical model to justify the results of the log-normal
simulations.

The structure of this paper is as follows. In Section~\ref{sec:method}
we briefly review the log-normal model, and then describe the log-normal
simulations and the analytical models. In Section~\ref{sec:results} we
show the main results of this paper, and then study the sparse sampling
strategy in Section~\ref{sec:sparse}. Section~\ref{sec:conclusion} is
devoted to conclusion and discussion. Throughout this paper we adopt the
concordance $\Lambda$CDM model, which is consistent with  the
WMAP 7-year results \citep{k11}. The model is characterized by
the  matter density $\Omega_{\rm m} =0.272$,
the baryon density $\Omega_{\rm b}=0.046$, the cosmological constant density
$\Omega_\Lambda=0.728$, the spectral index of the primordial power spectrum
$n_{\rm s}=0.97$, the present-day
rms mass density fluctuations $\sigma_8=0.81$, and the Hubble expansion
rate today $H_{0}=70.0~$km s$^{-1}$ Mpc$^{-1}$.

\section{Method}
\label{sec:method}

\subsection{Log-normal convergence field}
\label{sec:log}

To approximate the weak lensing convergence field for a $\Lambda$CDM model,
we employ the log-normal model.
The previous works based on ray-tracing simulations
 \citep[see][]{taru02,do06,hilb07,taka11b,neyr09,joach11,seo12} have shown 
that the log-normal model can serve as a fairly good approximation of the
 lensing field,
originating from the fact
that the three-dimensional matter field in large-scale
 structure
 is also
 approximated by the log-normal distribution
\cite[e.g.][]{cj91,kof94,kayo01}. 
The main reason of our use of the log-normal model is twofold;
it allows us to simulate many realizations of the lensing field without
running ray-tracing simulations 
as well as allows us to analytically compute statistical properties of
the lensing field including the non-Gaussian features. 

We assume that the lensing convergence field, $\kappa(\THE)$, 
 obeys the following one-point probability distribution function: 
\begin{eqnarray}
P(\kappa)
&=&  
 \frac{1}{\sqrt{2\upi}(\kappa/|\kappa_0|+1)\sigma_{\rm G}}
\nonumber\\
&&\hspace{-2em}\times
 \exp \left[ -\dfrac{ 
\left\{ |\kappa_0|
\ln \left( \kappa / |\kappa_0| + 1 \right)
 + \sigma_{\rm G}^2/(2|\kappa_0|) \right\}^2}{2\sigma_{\rm G}^2} \right],
\label{pdf_kappa}
\end{eqnarray}
%
for $\kappa > -|\kappa_0|$, and we set
$P(\kappa)=0$ for 
$\kappa \leq -|\kappa_0|$. Note that the above distribution satisfies 
$\int_{-|\kappa_0|}^\infty\!d\kappa~ P(\kappa)=1$
as well as $\int_{-|\kappa_0|}^\infty\!d\kappa~ \kappa P(\kappa)=0$.
The variance is $\sigma_\kappa^2 = \int_{-|\kappa_0|}^\infty\!d\kappa~ 
 \kappa^2 P(\kappa) = |\kappa_0|^2 [ \exp(\sigma_{\rm G}^2/|\kappa_0|^2) -1 ]$.
The log-normal distribution is specified by two parameters, 
$\sigma_{\rm G}$ and $\kappa_0$.
Throughout this paper, as for
$\kappa_0$, 
we use the empty beam value
for the fiducial cosmological model and the assumed source redshift
\citep{jain00}; 
$\kappa_0=-0.050$ for source redshift $z_s=0.9$.

Statistical properties of the log-normal
convergence field are {\em fully} characterized by
 the two-point correlation function of $\kappa(\THE)$ (see below).
What we meant by ``fully'' is any higher-order functions of
 the log-normal field are given as a function of products of the
 two-point function, but in a different form those 
of a Gaussian field.
As for the convergence power spectrum, we employ the model that
well reproduces the power spectrum seen in
ray-tracing simulations of a $\Lambda$CDM model
\citep[e.g.,][]{bs01}:
\BE
 C(\ell) = \frac{9 H_0^4 \Omega_{\rm m}^2}{4 c^4}
 \int_0^{r_{\rm s}}\!\! \rmd r 
 \frac{\left( r_{\rm s}-r  \right)^2}{a(r)^2 r_{\rm s}^2} 
 P_\delta\!\left(k=\frac{\ell}{r}; a(r) \right),
\label{cl}
\EE
where $r_{\rm s}$ is the comoving distance to the source,
 $a(r)$ is the scale factor at the distance $r$, and
$P_\delta(k;a)$ is the matter power spectrum given as a function of $k$
 and $a$. Note that we throughout this paper consider a single source
 redshift $z_s$ for simplicity; $z_s=0.9$.
In order to include effects of nonlinear gravitational clustering,
we use the revised version of halo-fit model \citep{smith03,taka12},
which can be analytically computed 
once the linear matter power spectrum and cosmological
model are specified. 
We employ the
fitting formula of \citet{eh99} to compute the input linear power
 spectrum. 

\subsection{Power spectrum and covariance estimation from 
the simulated log-normal lensing maps}
\label{2.2}

In order to estimate an expected measurement accuracy of the lensing
power spectrum against an assumed geometry of a hypothetical survey, we
use $1000$ simulation maps of the log-normal homogeneous and isotropic
convergence field.  

Following the method in \citet{neyr09} \citep[also
see][]{hilb11}, we generate the maps as follows.
(i) We choose the target power spectrum, $C(\ell)$, which the simulated
log-normal field is designed to obey. We employ the power spectrum
$C(\ell)$ expected 
for the assumed $\Lambda$CDM model and source redshift $z_s=0.9$, 
computed from Eq.~(\ref{cl}). The source redshift is chosen to 
 mimic the mean source redshift for a Subaru Hyper Suprime-Cam type survey.
For another parameter $\kappa_0$ needed to specify the log-normal model,  
we adopt the empty beam value in the cosmology,
$\kappa_0=-0.050$. Provided the target power spectrum
$C(\ell)$ and $\kappa_0$, we compute the power spectrum for the
corresponding Gaussian field, $C_{\rm G}(\ell)$, from the mapping
relation between the log-normal and Gaussian fields (see below).  (ii)
Using the Fast Fourier Transform (FFT) method, we generate a Gaussian
homogeneous and isotropic field, $\kappa_{\rm G}(\THE)$, from the power
spectrum $C_{\rm G}(\ell)$.  In making the map, we adopt $12180\times
12180 $ grids for an area of $203\times 203~{\rm deg}^2$ ($\simeq 4\upi$
steradian, i.e. all-sky area) so that the grid scale is 1 arcmin on a
side (because $203\times 60=12180$).  Since we used the FFT method,
the simulated Gaussian map obeys the periodic boundary condition; no
Fourier mode beyond the map size ($203~$deg.) exists.
The mean of $\kappa_{\rm G}$ is zero and the one-point 2nd-order moments is
 defined as $\sigma_{\rm G}^2\equiv \langle\kappa_{\rm G}(\THE)^2\rangle$
 (the variance of the FFT grid-based field).
(iii) We add the constant value, $-\sigma_{\rm G}^2/(2|\kappa_0|)$, to each
grid so that the mean of the Gaussian field becomes
$\ave{\kappa_{\rm G}(\THE)}=-\sigma_{\rm G}^2/(2|\kappa_0|)$.  
This constant shift is necessary so that the mean of the log-normal field
 is zero after the mapping (Eq.~\ref{mapping}).
(iv) Employing the log-normal mapping
\begin{equation}
\kappa(\THE) = |\kappa_0| \left[ \exp \left(
 \frac{\kappa_{\rm G}(\THE)}{|\kappa_0|} \right) -1 \right]
\label{mapping}
\end{equation}
we evaluate the log-normal field,
$\kappa(\THE)$, at each grid in the map. The variance of the
log-normal field is exactly related to that of the Gaussian field
$\kappa_{\rm G}(\THE)$ via
$\sigma_\kappa^2 = |\kappa_0|^2 [ \exp(\sigma_{\rm G}^2/|\kappa_0|^2) -1 ]$.
Since the grid size of 1 arcmin is still in the weak lensing regime,
$\sigma_\kappa^2\sim 10^{-4}$ and $\kappa_0=-0.05$, we can find
$\sigma_\kappa^2 \simeq \sigma_{\rm G}^2 + \sigma_{\rm G}^4/(2|\kappa_0|^2)$.  
The log-normal field $\kappa(\THE)$, simulated by this method, obeys the
one-point distribution given by Eq.~(\ref{pdf_kappa}).  Note that our
map-making method employs the flat-sky approximation even for the
all-sky area in order to have a sufficient statistics with the limited
number of map realizations as well as to include all the possible
super-survey modes beyond an assumed survey geometry. This assumption is
not essential for the following results, and just for convenience of our
discussion (see below for the justification).

The $n$-point correlation functions of the log-normal field
$\kappa(\THE)$ can be given in terms of the two-point correlation;
up to the four-point correlation functions are given as
\begin{eqnarray}
&&\hspace{-2em}\ave{\kappa(\THE_1)\kappa(\THE_2)}\equiv
\xi(|\THE_1-\THE_2|)=
|\kappa_0|^2\left[\eta_{12}-1\right],
\nonumber\\
&&\hspace{-2em}\ave{\kappa(\THE_1)\kappa(\THE_2)\kappa(\THE_3)}
=|\kappa_0|^3\left[
\eta_{12}\eta_{13}\eta_{23}-\eta_{12}-\eta_{13}-\eta_{23}+2
\right],
\nonumber\\
&&\hspace{-2em}
\ave{\kappa(\THE_1)\kappa(\THE_2)\kappa(\THE_3)\kappa(\THE_4)}=|\kappa_0|^4
\left[\eta_{12}\eta_{13}\eta_{14}\eta_{23}\eta_{24}\eta_{34}
\right.\nonumber\\
&&\hspace{2em}
\left.
-\eta_{12}\eta_{13}\eta_{23}-\eta_{12}\eta_{14}\eta_{24}
-\eta_{13}\eta_{14}\eta_{34}-\eta_{23}\eta_{24}\eta_{34}\right.\nonumber\\
&&\hspace{2em}\left.
+\eta_{12}+\eta_{13}+\eta_{14}+\eta_{23}+\eta_{24}+\eta_{34}-3
\right],
\label{eq:log_normal_4pt}
\end{eqnarray}
where 
\begin{equation}
 \eta_{12}\equiv \exp\left[\frac{\xi_G(|\THE_1-\THE_2|)}{|\kappa_0|^2}\right],
\end{equation}
and 
\begin{equation}
\xi_G(|\THE_1-\THE_2|)\equiv
\ave{\left[\kappa_G(\THE_1)+\frac{\sigma_G^2}{2|\kappa_0|}
\right]\left[\kappa_G(\THE_2)+\frac{\sigma_G^2}{2|\kappa_0|}
\right]}.
\end{equation}
Thus the log-normal field is, by definition, a non-Gaussian field and 
its higher-order moments are all non-vanishing. 
By using the relation $\eta_{12}=\xi/|\kappa_0|^2+1$, we can express all
the higher-order function in terms of the two-point function of the
log-normal field,
$\xi(\theta)$.

To include the effect of a survey geometry, we introduce 
the survey window function: 
 $W(\THE)=1$ if the angular position $\THE$ is 
inside the survey region, otherwise $W(\THE)=0$.
The total survey area is given as 
\begin{equation}
 \Omega_{W}\equiv \int\!\!\rmd^2\THE~W(\THE).
\label{survey_area}
\end{equation}
Then, the measured convergence field from a hypothetical survey
region is given by
$\kappa_W(\THE) = W(\THE) \kappa(\THE)$. 
For simplicity, 
we do not consider masking effects and any effects of incomplete
selection (e.g. inhomogeneous survey depth), 
which may be characterized by $W(\THE)<1$.

The Fourier-transform of the convergence field is 
\begin{equation}
 \tilde{\kappa}_W(\ELL) = \int\!\! \frac{\rmd^2 \ELL^\prime}
 {\left( 2\upi \right)^2} \tilde{W}(\ELL-\ELL^\prime)
 \tilde{\kappa}(\ELL^\prime), 
\end{equation}
Hereafter quantities with tilde symbol denote their Fourier-transformed
fields. 
Thus, via the window function convolution, the
Fourier field, $\tilde{\kappa}_W$, has contributions from modes
of length scales comparable with or beyond the survey size.

We use the FFT method to perform the discrete Fourier transform of the
simulated convergence field. Provided the above realizations of
the convergence field, suppose that
$\tilde{\kappa}_{{W}(r)}(\ELL)$ is the Fourier-transformed
field in the $r$-th realization map.
An estimator of the window-convolved power spectrum is defined as
\BE
 \hat{C}_{{W}(r)}(\ell) = 
\frac{1}{N_\ell}
 \sum_{|\ELL^\prime| \in \ell}
 \left| \tilde{\kappa}_{{W}(r)}(\ELL^\prime) \right|^2,
\label{eq:est_ps}
\EE
where the summation runs over Fourier modes satisfying the
 condition
$\ell - \Delta \ell/2 < |\ELL^\prime| < \ell + \Delta \ell/2$ 
($\Delta \ell$ is the bin width), and
 $N_\ell$ is the number of Fourier modes in the summation; 
$N_\ell\equiv \sum_{|\ELL^\prime| \in \ell}$.

We use the 1000 realizations to estimate the ensemble-average power spectrum:
\BE
  C_{W}(\ell) = \frac{1}{N_{r}} \sum_{r=1}^{N_{ r}} \hat{C}_{W(r)}(\ell),
\EE
where $N_{r}=1000$.
The power spectrum $C_W(\ell)$ differs from the underlying power
spectrum $C(\ell)$ due to the window convolution. Since the window
function can be exactly computed for a given survey geometry, we
throughout this paper consider $C_W(\ell)$ as an observable, and will not
consider any deconvolution issue. 

The covariance matrix of the power spectrum estimator describes an
expected accuracy of the power spectrum measurement for a given survey
as well as how the band powers of different multipole bins are
correlated with each other. Again 
we use the 1000 realizations to estimate the
covariance matrix:
\BEA
{\cal C}^W_{ij}
 && \hspace{-0.65cm} \equiv
{\rm Cov} \left[ C_W(\ell_i),C_W(\ell_j)
 \right]  \nonumber \\
 && \hspace{-4em} = \frac{1}{N_{ r}-1} \sum_{r=1}^{N_{ r}}
  \left[ \hat{C}_{W(r)}(\ell_i) - C_{W}(\ell_i) \right]
  \left[ \hat{C}_{W(r)}(\ell_j) - C_{W}(\ell_j) \right].
  \nonumber \\
\EEA
In this paper we consider up to 30 multipole bins for the power spectrum
estimation. 
We have checked that each covariance element
over the range of multipoles is well converged 
by using the 1000 realizations. 

The cumulative signal-to-noise ratio ($S/N$) of the power spectrum
measurement, integrated up to a certain maximum multipole $\ell_{\rm max}$,
is defined as
\BE
 \left( \frac{S}{N} \right)^2 = \sum_{\ell_i, \ell_j \le \ell_{\rm max}}
C_W(\ell_i)  [\bm{{\cal C}}^W]^{-1}_{ij} C_W(\ell_j),
\label{sn}
\EE
where $[\bm{{\cal C}}^W]^{-1}$ is the inverse of the covariance matrix. 

\subsection{Analytical model of the power spectrum covariance
  including the super-sample covariance}
\label{sec:analytical}

In this section, we follow the formulation in \citet{th13} \citep[see
also][]{Lietal:14} to analytically derive the power spectrum covariance
for the log-normal field, including the super-sample covariance (SSC)
contribution. We will then use the analytical prediction to compare with the
simulation results.

The window-convolved power spectrum is expressed in terms of the
underlying true power spectrum as
\BE
 C_W(\ell) = \frac{1}{\Omega_{W}} \int_{|\ELL^\prime| \in \ell}\!\!
 \frac{\rmd^2 \ELL^\prime}{A_\ell} \int\!\! \frac{\rmd^2 \bq}{\left( 2\upi \right)^2}
 \left| \tilde{W}(\bq) \right|^2 C(\ELL'-\bq),
\label{eq:cl_W}
\EE
where $A_\ell$ is the Fourier-space area of the integration range of
$\rmd^2\ELL^\prime$:  $A_\ell\equiv
\int_{|\ELL^\prime|\in\ell}\!\!\rmd^2\ELL^\prime$. 
Note that here and hereafter we use the vector notation $\bq$,
instead of $\ELL$, to denote super-survey modes with $q\ll l$ for
presentation clarity.

The covariance matrix is given as
\BE
{\cal C}^W_{ij} = \frac{2}{N_{\ell_i}} C_W(\ell_i)^2\delta^K_{ij}
 +\bar{T}^W\!(\ell_i,\ell_j),
\label{eq:cov_def}
\EE
where $\delta^K_{ij}$ is the Kronecker delta function; $\delta^K_{ij}=1$
if $\ell_i=\ell_j$ to within the bin width, otherwise $\delta^K_{ij}=0$.
The first term is the Gaussian covariance contribution, which has
only the diagonal components; in other words, it ensures that
the power spectra of different bins are independent.
The second term, proportional to $\bar{T}^W(\ell_i,\ell_j)$, is
the non-Gaussian contribution arising from the connected part of 4-point
correlation function, i.e. trispectrum in Fourier space.
The trispectrum contribution is given in terms of the underlying 
true trispectrum, convolved with the survey window function, as
\begin{eqnarray}
\bar{T}^W\!\! (\ell_i,\ell_j) &=&  \frac{1}{\Omega_{W}}
\int_{|\ELL|\in \ell_i}\!\!\frac{\rmd^2 \ELL}{A_{\ell_i}}
\int_{|\ELL^\prime|\in \ell_j}\!\!\frac{\rmd^2 \ELL^\prime}{A_{\ell_j}}
\nonumber\\
&&\hspace{-3em}\times
\left[\int\!\prod_{a=1}^4\frac{\rmd^2\bq_a}{(2\upi)^2}\tilde{W}(\bq_a)\right] 
(2\upi)^2\delta_D^2(\bq_{1234})
 \nonumber \\
&&\hspace{-3em}
 \times 
 T(\ELL+\bq_1,-\ELL+\bq_2,\ELL^\prime+\bq_3,-\ELL^\prime+\bq_4),
\label{eq:4pt_cov}
\end{eqnarray}
where 
$\bq_{1 \cdots n}\equiv \bq_1+ \cdots +\bq_n$,
$\delta_D^2(\bq)$ is the Dirac delta function,
 and $T$ is the true trispectrum.
The convolution with the window function means that different 4-point
configurations separated by less than the Fourier width of the window
function account for contributions arising from super-survey modes.

Using the change of variables 
$\ELL+\bm{q}_1 \leftrightarrow \ELL$ and
 $\bm{q}_1+\bm{q}_2\leftrightarrow \bm{q}$
under the delta function and the approximation
 $\ell_i,\ell_j \gg q$, one can find that the
non-Gaussian covariance term arises from the following squeezed
quadrilaterals where two pairs of sides are nearly equal and opposite:
\begin{equation}
 T(\ELL,-\ELL+\bq,\ELL^\prime,-\ELL^\prime-\bq).
\end{equation}
For the log-normal field, we can analytically compute the 4-point
function as explicitly given in Appendix~\ref{app:4pt}. Plugging the
above squeezed trispectrum into Eq.~(\ref{trisp}) yields
\begin{equation}
 T(\ELL,-\ELL+\bq,\ELL^\prime,-\ELL^\prime-\bq) \simeq
 T(\ELL,-\ELL,\ELL^\prime,-\ELL^\prime) +
 \frac{4}{\kappa_0^2}C(q)C(\ell)C(\ell^\prime), 
\label{eq:4pt_squeezed}
\end{equation}
where the first term $T(\ELL,-\ELL,\ELL^\prime,-\ELL^\prime)$ arises
from the sub-survey
modes 
and is given
in terms of products of the power spectrum (see Eq.~\ref{trisp}).
For the above equation, we ignored the higher-order terms of
$O(C^4/|\kappa_0|^4)$, based on the fact $\xi/|\kappa_0|^2\ll 1$ as we
discussed around Eq.~(\ref{eq:log_normal_4pt}).
The 2nd term describes extra correlations between the modes $\ELL$ and
 $\ELL^\prime$ via super-survey modes $C(q)$ with $q\ll \ell,\ell^\prime$.

Hence, by inserting Eq.~(\ref{eq:4pt_squeezed}) into
Eq.~(\ref{eq:4pt_cov}), we can find that 
the power spectrum covariance for the log-normal field is given as
\begin{equation}
 {\cal C}^W_{ij}\simeq {\cal C}^{G}_{ij} + {\cal C}^{T0}_{ij} 
+ {\cal C}^{\rm SSC}_{ij},
\label{eq:cov_unified}
\end{equation}
where
\begin{eqnarray}
{\cal C}^{\rm G}_{ij} && \hspace{-0.6cm} \equiv \frac{2}{N_{\ell_i}}
 C_W(\ell_i)^2 \delta^K_{ij}, \\
{\cal C}_{ij}^{\rm T0} && \hspace{-0.6cm} \equiv \frac{1}{\Omega_W}
\int_{|\ELL|\in \ell_i}\!\!\frac{\rmd^2\ELL}{A_{\ell_i}}
\int_{|\ELL^\prime|\in \ell_j}\!\!\frac{\rmd^2\ELL^\prime}{A_{\ell_j}}
T(\ELL,-\ELL,\ELL^\prime,-\ELL^\prime), \\
{\cal C}_{ij}^{\rm SSC}&& \hspace{-0.6cm} \equiv 
\frac{4}{\kappa_0^2}
 (\sigma_W)^2 C(\ell_i)C(\ell_j), 
\end{eqnarray}
with
\begin{eqnarray}
(\sigma_W)^2 && \hspace{-0.6cm} \equiv
 \frac{1}{\Omega_W^2}\int\!\frac{\rmd^2 \bq}{(2\upi)^2}|\tilde{W}(\bq)|^2
 C(q).
\label{sigma_w}
\end{eqnarray}
The first and second terms on the r.h.s. of
Eq.~(\ref{eq:cov_unified}) are standard covariance terms, as originally
derived in \citet{szh99}, and arise from the sub-survey modes.
The third term is the SSC term. It scales
with the survey area through $(\sigma_W)^2$, while the standard
terms scale with $1/\Omega_W$.

Eq.~(\ref{sigma_w}) is rewritten as $(\sigma_W)^2=\ave{\bar{\kappa}_W^2}$,
where $\bar{\kappa}_W$ is the mean
convergence averaged within the survey region, defined as  
$\bar{\kappa}_W\equiv (1/\Omega_W)\int\!\!d^2\THE~W(\THE)\kappa(\THE)$.
Thus $(\sigma_W)^2$ can be realized as the variance of the background
convergence mode or the mean density mode across the survey area.
The variance $(\sigma_W)^2$ is the key quantity to understand the
effect of the power spectrum covariance on survey geometry as we will
show below. If we consider a sufficiently wide-area survey, 
$(\sigma_W)^2$ arises from the convergence field in the linear regime.
Thus the variance $(\sigma_W)^2$ can be easily computed
for any survey geometry, either by evaluating Eq.~(\ref{sigma_w})
directly, or using Gaussian realizations of the linear convergence field.
For convenience of
 the following discussion, we also give another expression of
$(\sigma_W)^2$ in terms of the two-point correlation function as
\BE
 \left( \sigma_W \right)^2 = \frac{1}{\Omega_W^2} \int\!\!\rmd^2\THE
 \int\!\! \rmd^2 \THE^\prime ~W(\THE) W(\THE^\prime) \xi(|\THE-\THE^\prime|).
\label{xi}
\EE

As discussed in \citet{th13}, we can realize that the SSC term is
 characterized by the response of $C(\ell)$ to a fluctuation in the
 background density mode $\bar{\kappa}_W$:
\begin{equation}
 {\cal C}^{\rm SSC}_{ij}=(\sigma_W)^2\frac{\partial C(\ell_i)}{\partial
\bar{\kappa}_W}
\frac{\partial C(\ell_j)}{\partial \bar{\kappa}_W}.
\end{equation}
For the log-normal field, the power spectrum
response is found to be
\begin{equation}
 \frac{\partial C(\ell)}{\partial
  \bar{\kappa}_W}=\frac{2}{\kappa_0}C(\ell).
\label{eq:ps_response}
\end{equation}
In this approach we approximated the super-survey modes in each
survey realization to be represented by the mean density fluctuation
$\bar{\kappa}_W$. In other words we ignored the high-order super-survey
modes such as the gradient and tidal fields, which have scale-dependent
variations across a survey region. We will below test the accuracy of
this approximation.

Here we also comment on the accuracy of the flat-sky approximation.
Let us first compare the convergence power spectra computed in the flat-
 and all-sky approaches.
We used the formula in  \cite{Hu2000} (Eqs.~28 and A11 in the paper)
to evaluate the all-sky power spectrum for the fiducial $\Lambda$CDM model.
We found that the flat-sky power spectrum is smaller than the all-sky
spectrum in the amplitude
 by 30, 13 and 7 and 4\% at low multipoles $\ell=1, 2$, 3 and 4,
 respectively. 
The
 relative difference becomes increasingly smaller by less 
than $2\%$ at the higher multipoles 
$\ell \geq 5$.
For the linear variance 
$(\sigma_W)^2$ in the all-sky approach
we can compute it as
$(\sigma_W)^2=(1/\Omega_S^2)\sum_\ell (2\ell+1) |\tilde{W}(\ell)|^2 C(\ell)$, 
where the window function 
$\tilde{W}(\ell)$ and the power spectrum $C(\ell)$ need to be computed
in harmonic space  \citep[e.g.,][]{mhb14}. 
We used the HEALPix software \citep{Gorski05} to evaluate
$\tilde{W}(\ell)$ for a given survey geometry such as rectangular shaped
geometries we will consider below.
We found that the flat-sky variance agrees with the full-sky variance to
within $1.2\%$ for the rectangular geometries.  Thus we conclude that,
since we are interested in the effect of SSC on the power spectrum at
high multipoles in the nonlinear regime, an inaccuracy of the flat-sky
approximation is negligible and
 does not change the results we will show below.

\subsection{Test of analytical model with simulations}
\label{csr}

\begin{figure}
\begin{center}
\includegraphics[width=8cm]{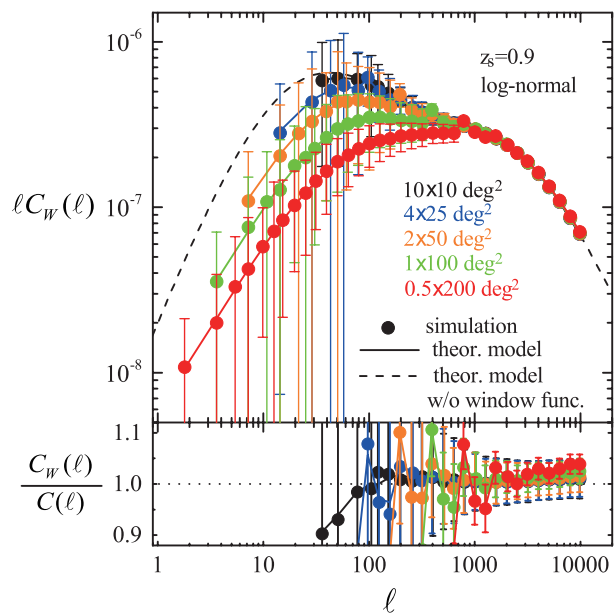}
\caption{The window-convolved power spectra of the log-normal lensing
field for different survey geometries, but keeping the area fixed to 100
deg$^2$. As denoted by legend, the circle points are the average power
 spectra from the 1000 simulation maps (see Section~\ref{2.2}), for
geometries of $10\times 10$ (black points), $4\times 25$ (blue),
$2\times 50$ (orange), $1\times 100$ (green) and $0.5\times 200$
deg$^2$ (red), from top to bottom points at $\ell\simeq 50$. The error bar
around each point denotes $\pm 1\sigma$ scatters of the 1000
realizations. For illustrative purpose, we here plot $\ell C_W(\ell)$,
making the power spectra amplitude relatively scale-independent over a
range of $\ell=[1,10^4]$. For comparison, the dashed curve shows the
underlying true spectrum without the window function convolution. The
solid curve around each point shows the analytical prediction, computed
from Eq.~(\ref{eq:cl_W}). The lower panel shows the fractional
difference of each power spectrum compared to the true power spectrum,
where the $y$-axis plotted range is chosen to illuminate the difference
in the range of $\ell>100$.  \label{fig:clw}}
\end{center}
\end{figure}
In this subsection, we test the analytical model of the power
spectrum covariance 
against the simulation 
of the 1000 convergence maps in Section~\ref{2.2}.

Before going to the comparison, Fig.~\ref{fig:clw} shows the
window-convolved power spectra for different survey geometries, with the
area being fixed to $100$ sq. degrees.  We consider a square shape
($10\times 10$~ deg$^2$) and rectangular shaped geometries with various
side length ratios; $4\times 25$, $2\times 50$, $1\times 100$ and
$0.5\times 200~$deg$^2$, respectively. For the discrete Fourier
decomposition, we apply FFT to the rectangular shaped region where
$W(\THE)=1$. The different geometries thus have different Fourier
resolution as follows. Let us denote the survey geometry as
$\Omega_W=a\times b$, where $a$ (radian) is the longer side length and $b$
(radian) is the shorter side; e.g., $a=100\times \upi/180=1.75$ rad and
$b=0.0175$ rad for the case of $1\times 100$ deg$^2$. Thus the
fundamental Fourier mode is $\ell_f=2\upi/a$ or $2\upi/b$ along the $a$-
or $b$-direction, respectively, meaning a finer Fourier resolution along
the $a$-direction. However, since all the simulated maps have the same
grid scale of $1$ arcmin, the Nyquist frequency (the
maximum multipole probed) is the same, $\ell_{\rm Ny}=\upi/1~{\rm
arcmin}=10800$, for all the survey geometries.
The window convolution mixes different Fourier modes,
causing extra correlations between different bins.
As can be found from
Fig.~\ref{fig:clw}, the convolution causes a significant
change in the convolved power spectrum compared to the underlying true spectrum
at multipoles $\ell\simlt 2\upi/b$. The change is more significant and
appears up
to higher multipoles for a more elongated survey geometry, due to a
greater mixture of different Fourier modes. 
At larger multipole bins $\ell\simgt 2\upi/b$, the window function stays
constant within the multipole bin and 
all the convolved power spectra appear similar to 
each other to within 5 per cent. 
The solid curves are the analytical predictions computed from
Eq.~(\ref{eq:cl_W}). Thus 
the convolved power spectrum can be analytically computed
 if the window function is known.
As can be found from the lower panel, the scatter around each point,
computed from the 1000 realizations, is smaller for a more elongated
survey geometry, as we will further study below.

\begin{figure}
\begin{center}
\includegraphics[width=11.5cm]{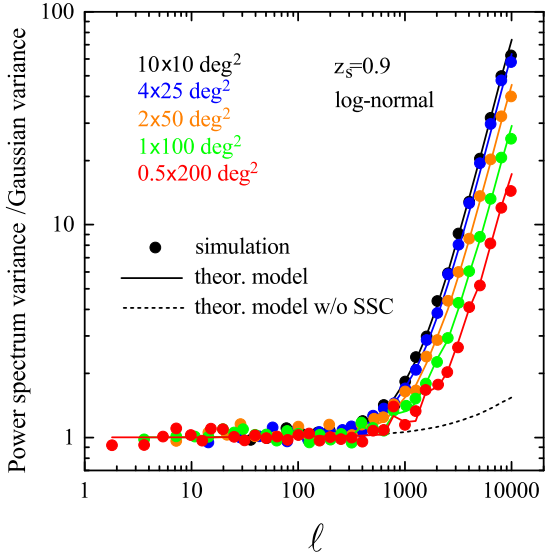}
\caption{The diagonal elements of the power spectrum covariance for the
log-normal convergence field, for different survey geometries as in
Fig.~\ref{fig:clw}.  Here we plot the diagonal elements relative to the
Gaussian covariance, ${\cal C}^W(\ell,\ell)/[2C_W(\ell)^2/N_{\ell}]$; 
a deviation from
unity is due to the non-Gaussian covariance contribution.
The symbols are the simulation results, while the solid curves are the
analytical predictions computed from Eq.(\ref{eq:cov_unified}), which
show a remarkably nice agreement with the simulation results. The dashed
curve is the analytical prediction without the super sample covariance
(SSC) contribution (the third term in Eq.~\ref{eq:cov_unified}).
Thus the non-Gaussian covariance contribution is mainly from the SSC
effect.} \label{cl_variance}
\end{center}
\end{figure}
\begin{table}
\begin{center}
\begin{tabular}{cc}
\hline
  survey geometry   &  $(\sigma_W)^2$  \\
\hline
 $10 \times 10 ~{\rm deg}^2$  &  $8.7 \times 10^{-7}$  \\
 $4 \times 25 ~{\rm deg}^2$ & $7.3 \times 10^{-7}$  \\
 $2 \times 50 ~{\rm deg}^2$ & $5.2 \times 10^{-7}$  \\
 $1 \times 100 ~{\rm deg}^2$ & $3.3 \times 10^{-7}$  \\
 $0.5 \times 200 ~{\rm deg}^2$ & $1.9 \times 10^{-7}$  \\
\hline
\end{tabular}
\caption{The variance of the background convergence mode, $(\sigma_W)^2$
(Eq.~\ref{sigma_w}), for different rectangular geometries, with a fixed
survey area of 100 sq. degrees as in Fig.~\ref{fig:clw}. The more
 elongated geometry has the smaller $(\sigma_W)^2$ for the lensing
power spectrum of $\Lambda$CDM model.
}
\label{table_sigmaw}
\end{center}
\end{table}
In Fig.\ref{cl_variance}, we study the diagonal components of the
window-convolved
power spectrum covariance as a function of the multipole bins, for
different survey geometries as in Fig.~\ref{fig:clw}.
Here we plot the diagonal covariance components
relative to the Gaussian covariance (the first term of
Eq.~\ref{eq:cov_unified}). Hence when the curve deviates from unity in the
$y$-axis, it is from the non-Gaussian covariance contribution (the 2nd
and 3rd terms in Eq.~\ref{eq:cov_unified}). The log-normal model
predicts significant non-Gaussian contributions at $\ell \simgt$ a few
$10^2$.  Although the relative importance of the non-Gaussian covariance
depends on the bin width on which the Gaussian covariance term depends
via $N_{\ell_i}(\propto 2\upi \ell_i \Delta \ell)$, an amount of the
 non-Gaussian contribution in the log-normal model is indeed similar to
 that seen from the ray-tracing simulations in \citet{sato09} as we will
 again discuss later. 

Fig.~\ref{cl_variance} shows a significant difference for
different survey geometries at $\ell\simgt $ a few $10^2$. Recalling
that the
window-convolved spectra for different geometries are similar at
$\ell\simgt $a few $10^2$ as shown in Fig.~\ref{fig:clw}, 
we can find that
the difference is due to the different SSC contributions, because
the survey geometry dependence arises mainly from the SSC term via
$(\sigma_W)^2$ in Eq.~(\ref{eq:cov_unified}). 
The most elongated rectangular geometry of $0.5 \times 200$ deg$^2$
 shows a factor $4$ smaller covariance 
amplitude than the square-shaped geometry of $10\times 10$ deg$^2$, 
the most compact geometry among the 5 geometries considered here.   
This can be confirmed by the analytical model of the
power spectrum covariance; the solid curves,
computed based on Eq.~(\ref{eq:cov_unified}), show remarkably nice
agreement with the simulation results
\footnote{Exactly speaking, for the analytical
predictions of the covariance, we used the
window-convolved power spectra, computed from Eq.~(\ref{eq:cl_W}), instead
of the true power spectra in order to
compute products of the power spectra appearing in 
the covariance terms ${\cal C}^{\rm G}$ and ${\cal C}^{\rm T0}$. This
gives about 5--20\% improvement in the agreement with the simulation
results of different geometries at low multipoles $\ell\simlt 2\upi/b$. 
Note that this treatment does not cause any difference  
at the higher multipole bins.}.
Table~\ref{table_sigmaw} 
clearly shows that $(\sigma_W)^2$ for the
elongated rectangular geometry of $0.5 \times 200$ deg$^2$ is about factor 4
smaller than that for the square-shaped geometry of $10\times 10$
deg$^2$, which explains the relative differences 
in Fig.~\ref{cl_variance}. For
comparison, the dotted curve in the figure shows the analytical
prediction if the SSC term, ${\cal C}^{\rm SSC}_{ii}$, is ignored 
(in this case no difference for different geometries). 
The analytical model without the SSC term
significantly underestimates
the simulation results at the nonlinear scales.

\begin{figure}
\begin{center}
\includegraphics[width=8cm]{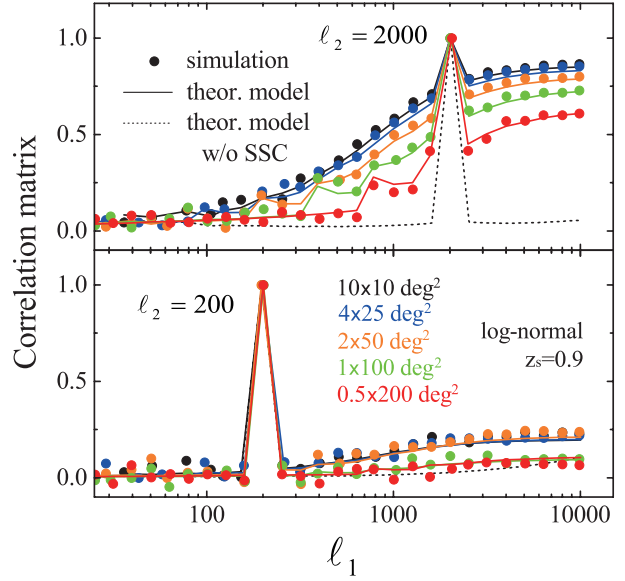}
\caption{The off-diagonal elements of the power spectrum covariance
matrix, for the different survey geometries as in the previous figure.
Here we plot the correlation coefficient matrix, $r(\ell_1,\ell_2)$
(Eq.~\ref{eq:r_ij}) as a function of $\ell_1$ for $\ell_2=2000$ 
(upper panel) or $200$ (lower).  The symbols are the simulation
results, while the solid curves are the analytical predictions
(Eq.~\ref{eq:cov_unified}); the two are in nice agreement with each other.
The dotted curve is the analytical prediction without the SSC effect. 
}
\label{fig_corrmat}
\end{center}
\end{figure}
Fig.\ref{fig_corrmat} shows the off-diagonal elements of the covariance
 matrix. For illustrative purpose, we study the correlation coefficients
 defined as 
\BE
r(\ell_1, \ell_2) \equiv \frac{{\cal C}^W(\ell_1,\ell_2)}
{\sqrt{{\cal C}^W(\ell_1,\ell_1){\cal C}^W(\ell_2,\ell_2)}}.
\label{eq:r_ij}
\EE
The correlation coefficients are normalized so that 
$r(\ell_1,\ell_2)=1$ for the diagonal
components with  $\ell_1=\ell_2$. 
For the off-diagonal components with $\ell_1\ne \ell_2$, $r\rightarrow
1$ implies strong correlation between the power spectra of the two
bins, while $r=0$ corresponds to no correlation.
The figure again shows a significant correlation between the different
multipole bins for $\ell_2=2000$, due to the significant SSC contribution. 
Similarly to Fig.~\ref{cl_variance}, 
the analytical model nicely reproduces the simulation results over the
range of multipoles and for the different geometries. 
For comparison, the dotted curve shows the prediction
without the SSC term.
As clearly seen from the figure, the correlation is smaller for more elongated
 survey geometry, due to the smaller $(\sigma_W)^2$ (see
 Table~\ref{table_sigmaw}).

\begin{figure}
 \begin{center}
  \includegraphics[width=8cm]{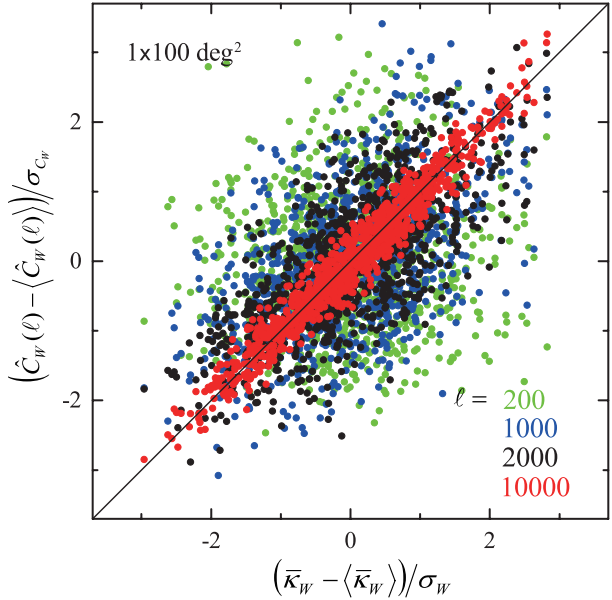}
 \end{center}
\caption{Scatters between the band power of the power spectrum at each
 multipole bin, $\hat{C}_W(\ell)$, and the mean convergence of the survey
 region, $\bar{\kappa}_W$, in the 1000 realizations, for the rectangular
 survey geometry of $1\times 100$ deg$^2$ in Fig.~\ref{fig:clw}. For the
 higher multipoles $\ell\ge 1000$, the two scatters display a tight
 relation well approximated by
$\left[ \hat{C}_W(\ell)-\ave{\hat{C}_W(\ell)} \right]/\sigma_{C_W}=
 \left[ \bar{\kappa}_W-\ave{\bar{\kappa}_W} \right]/\sigma_W$, where
 $\sigma_{C_W}$ and $\sigma_W$ are the variances computed from
 the same $1000$ realizations. Note $\ave{\bar{\kappa}_W}\simeq 0$.
 \label{fig:cl_scatter}}
\end{figure}
As we mentioned below Eq.~(\ref{eq:ps_response}), the
approximation we used for the analytical model of the SSC effect is that
we modeled the super-survey modes by the mean density fluctuation in
each survey realization, $\bar{\kappa}_W$. To test the validity of this
approximation, in Fig.~\ref{fig:cl_scatter} we study how a scatter of
the power spectrum estimation in each realization, $\hat{C}_W(\ell)$, is
correlated with the mean density in the realization,
$\bar{\kappa}_W$. Here we used the 1000 realizations for the rectangular
geometry of $1\times 100$ deg$^2$ as in Fig.~\ref{fig:clw}, but checked
that the results are similar for other geometries.  For the higher
multipoles in the nonlinear regime, $\ell\ge 1000$, the scatters of the
two quantities display a tight correlation reflecting the fact that the
mean density fluctuation is a main source of the scatters of the band
power on each realization basis.  In other words, the higher-order
super-survey modes such as the gradient and tidal fields that have
scale-dependent variations across the survey region are not a
significant source of the scatter in the power spectrum; also see Fig.~6
in \citet{Lietal:14} and Figs.~2 and 3 in \citet{Lietal:14b} for the
similar discussion. We have also checked that the averaged relation of
the scatters is well described by the power spectrum response as implied
by Eq.~(\ref{eq:ps_response}): $\left[ \hat{C}_W(\ell)-\ave{\hat{C}_W(\ell)}
 \right] \simeq (\partial C_W(\ell)/\partial \bar{\kappa}_W) \bar{\kappa}_W=
(2/\kappa_0)C_W(\ell)\bar{\kappa}_W$. The tight relation is probably due
to the fact that the power spectrum at a given multipole bin $\ell$ is
estimated from the angle average of the Fourier coefficients
$|\tilde{\kappa_{\ELL}}|^2$ with the fixed length $|\ELL|$ and therefore
is sensitive to the angle-averaged super-survey modes, i.e. the mean
density fluctuation on each realization basis. With this result, an
optimal survey geometry or strategy for mitigating the SSC contamination
can be studied by monitoring the mean density field $\bar{\kappa}_W$ or
the variance $\sigma_W^2$ against survey geometry, and therefore the
optimal survey geometry we will show below is valid even on each
realization basis.  We would like to note that the higher-order
correlation function such as the bispectrum may display a sensitivity to
the higher-order super-survey modes. This is beyond the scope of this
paper, and needs to be further studied.

\begin{figure}
\begin{center}
\includegraphics[width=8cm]{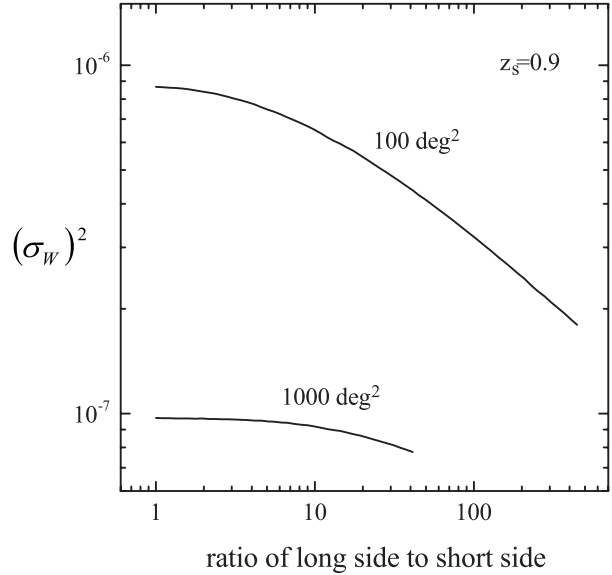}
\caption{The variance of the background convergence mode, $(\sigma_W)^2$,
as a function of the side length ratio for rectangular survey geometry
of 100 or 1000 sq. degrees. The more elongated geometry (the greater
ratio) has the smaller $(\sigma_W)^2$.  } \label{fig_sigmaw}
\end{center}
\end{figure}
Does the more elongated geometry for a fixed area always have the
smaller SSC contribution? The answer is yes for a continuous
survey geometry, as can be found from Fig.~\ref{fig_sigmaw}. For the
lensing field expected for a $\Lambda$CDM model, the variance of the
background convergence modes, $(\sigma_W)^2$, becomes smaller for the
more elongated geometry. 
For  a statistically isotropic and homogeneous field, the impact of the
non-Gaussian covariance can be mitigated, as long as Fourier modes along
the longer side length direction can be sampled, even if the modes the
shorter side direction is totally missed. 
This conclusion is perhaps counter-intuitive, 
but this is a consequence of non-Gaussian features in non-linear structure
 formation of a $\Lambda$CDM model.

\section{Results}
\label{sec:results}

In this section, we study the impact of different survey geometries on
the lensing power spectrum measurement, using the simulated maps of
log-normal lensing field. In studying this, we do not consider any
observational effect: intrinsic shape noise and imperfect shape
measurement error. We focus on the effect of survey geometry for clarity
of presentation.

\subsection{Signal-to-Noise Ratio}

\begin{figure}
\begin{center}
\includegraphics[width=11.5cm]{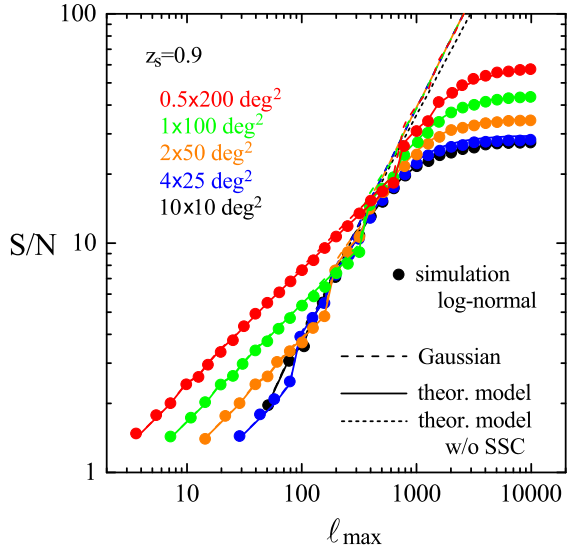} 
\caption{ The cumulative signal-to-noise ratio ($S/N$) of the power
spectrum measurement, integrated up to a certain maximum multipole
($\ell_{\rm max}$), for different survey geometries as in
Fig.~\ref{fig:clw}.
The minimum multipole for the different geometries is taken from the
fundamental Fourier mode available from each geometry. 
The symbols show the simulation results; the most elongated geometry
gives the highest $S/N$ values over the range of multipoles we consider.
The dashed curves show the $S/N$ value expected in a Gaussian field for
each geometry, which we estimated by counting the number of Fourier
 modes around each multipole bin (see text for details). 
For comparison, the solid and dotted curves show the analytical
prediction with and without the SSC effect.
The solid curves overlap the dashed curves in the linear regime
 ($\ell <$ a few $10^2$).}\label{sn_rectang}
\end{center}
\end{figure}
Fig.~\ref{sn_rectang} shows the cumulative signal-to-noise ratio ($S/N$)
of the window-convolved power spectrum, integrated up to a certain
maximum multipole $\ell_{\rm max}$ (Eq.~\ref{sn}), for different survey
geometries studied in Fig.~\ref{fig:clw}. The $S/N$ is independent of
the bin width and quantifies the total information content inherent in
the power spectrum measurement taking into account cross-correlations
between the different multipole bins. For the minimum multipole
$\ell_{\rm min}$, we adopt the
fundamental mode of a given survey geometry ($\ell_{\rm min}=2\upi/a$ as
we discussed above).
The inverse of $S/N$ gives the fractional error of estimation of
the power spectrum
amplitude parameter when using the power spectrum information up to
$\ell_{\rm max}$ for a given survey, assuming that the shape of the
power spectrum is perfectly known.  Fig.~\ref{sn_rectang} clearly shows
that the $S/N$  significantly 
varies with different survey geometries, over the range of multipoles.
To understand the results, again let us 
denote the geometry as $\Omega_{W}=a\times b$
($a$ is the longer side length as before). For the range
of multipole bins, $2\upi/a\simlt \ell\simlt 2\upi/b$, only Fourier modes
along the $a$-direction are sampled, therefore this regime is 
one-dimensional, rather than two-dimensional.
Hence, when measuring the power spectrum around a certain
$\ell$-bin with the bin width $\Delta \ell$, the number of the sampled
modes is given as $N_{\ell}\simeq \Delta\ell/(2\upi/a)$. For the bins
$\ell\simgt \upi/b$, the Fourier modes in the two-dimensional space
can be sampled. Hence the number of modes around the $\ell$-bin,
$N_\ell\simeq 2\upi \ell\Delta\ell/[(2\upi)^2/(ab)]=
2\upi\ell\Delta\ell/[(2\upi)^2/\Omega_W]$. The dashed curves give the $S/N$
values expected for a Gaussian field, estimated by accounting for the
number of Fourier modes for each survey geometry. To be more precise,
the Gaussian covariance is given by ${\cal
C}^G(\ell,\ell)=2C_W(\ell)^2/N_\ell$ and therefore
$(S/N)^2=\sum_{\ell}^{\ell_{\rm max}}N_\ell/2$. Since we adopt the
logarithmically-spaced bins of $\ell_{\rm max}$, the Gaussian
predictions $(S/N)^2\propto \ell_{\rm max}$ at $\ell_{\rm max}\simlt
2\upi/b$, while $(S/N)^2\propto \ell_{\rm max}^2$ at $\ell_{\rm
max}\simgt 2\upi/b$. 
The Gaussian prediction shows a nice agreement with the
simulation results in the linear regime, $\ell_{\rm max}\simlt $ a few $10^2$. 
In the linear regime, the figure
shows a greater $(S/N)^2$ for a more elongated geometry due to the larger
$N_\ell$. It is also worth noting that the elongated geometry allows for
an access to the larger angular scales (i.e. the lower multipoles).

For the regime of large multipoles, $\ell_{\rm max}>$ a few $10^2$, the
non-Gaussian covariance significantly
degrades the information content compared to the Gaussian expectation.
The $S/N$ value does not increase at
$\ell \simgt $a few $10^3$, implying that the power spectrum can
not extract all the information in the log-normal field, i.e. the
Gaussian information content from which the log-normal map is generated. 
The degradation is mainly due to the SSC effect, 
as shown by the dotted curve (also see
Figs.~\ref{cl_variance} and \ref{fig_corrmat}).
The more elongated survey geometry mitigates the SSC effect; the most
elongated survey geometry of $0.5 \times 200$ deg$^2$ gives about factor
2 higher $S/N$ than in the most compact square geometry of
$10\times 10$ deg$^2$. 
Although we here used the 1000 realizations to compute the $S/N$
values, we have checked that the analytical model in
Section~\ref{sec:analytical} can reproduce the simulation results.

We have so far used the log-normal convergence field, as an
approximated working example of the nonlinear large-scale structure for
a $\Lambda$CDM model. In Appendix~\ref{app:sato}, 
using the 1000 realizations of ray-tracing
simulations in \citet{sato09}, each of which has much smaller area
($25~$deg$^2$) and was built based on $N$-body simulations of
$\Lambda$CDM model, we
also found that an elongated survey geometry gives a larger $S/N$
value compared to a square shape, although the geometry size 
is limited to a
much smaller area, $\Omega_W\simeq 0.39~$deg$^2$, due to the available
area sampled from the ray-tracing simulation area ($25$ deg$^2$).

\subsection{An implication for cosmological parameter estimation}

\begin{figure}
\begin{center}
\includegraphics[width=8cm]{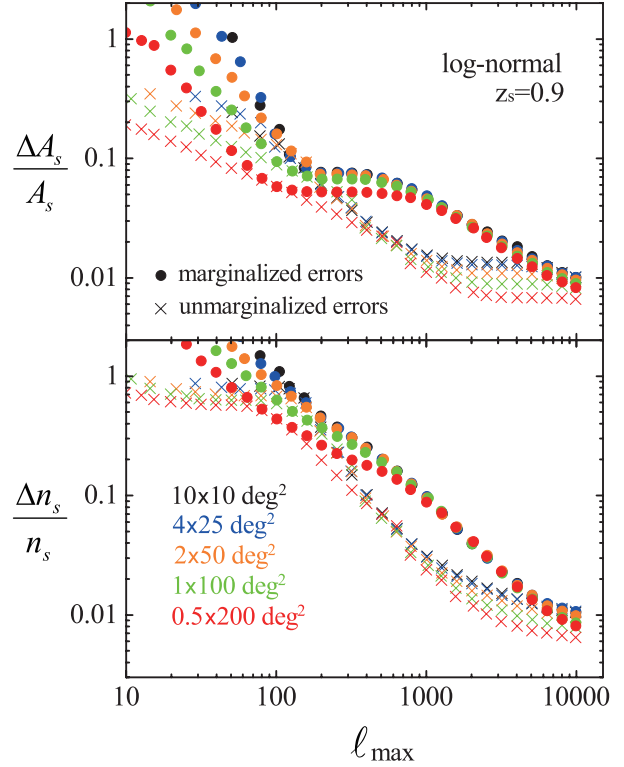}
\caption{The fractional error of cosmological parameters, the
 primordial power spectrum amplitude $A_s$ (upper panel) and the
 spectral tilt parameter $n_s$ (lower), expected from the power spectrum
 measurement for different survey geometries, but for a fixed survey
 area of $100$ sq. degrees (see text for the details).
Note that the other cosmological parameters are fixed to their fiducial
 values. 
The error is shown as a function of the maximum multipole $\ell_{\rm
 max}$ up to which the power spectrum information is included in the
 parameter forecast.
The filled circle symbols are the marginalized errors, while 
the cross symbols denote the unmarginalized errors, the error when another
 parameter (another $n_s$ or $A_s$) is fixed to the fiducial value.}
\label{fig_fisher}
\end{center}
\end{figure}
What is the impact of survey geometry on cosmological parameter estimation? 
Can we achieve a higher precision of cosmological parameters by just taking
an optimal survey geometry, for a fixed area (although we here consider
a continuous survey geometry)?  The SSC causes correlated up- or
down-scatters in the power spectrum amplitudes over a wide range of
multipole bins. The correlated scatters to some extent preserve a shape
of the power spectrum, compared to random scatters over different
bins. Hence, the SSC is likely to most affect parameters that are
sensitive to the power spectrum amplitude, e.g. the primordial curvature
perturbation $A_s$. On the other hand, other parameters that are
sensitive to the shape, e.g. the spectral tilt of the primordial power
spectrum $n_s$, is less affected by the SSC \citep[see also][for the
similar discussion]{TakadaJain:09,Lietal:14b}.

Based on this motivation, we use the simulated log-normal convergence
maps to estimate an expected accuracy of the parameters ($A_s$, $n_s$) as a
function of different survey geometries, using the Fisher information
matrix formalism. 
When including the power spectrum information up to a
certain maximum multipole $\ell_{\rm max}$, the Fisher matrix for the
two parameters is given as
\BE
 F_{ab}(\LAMBDA) = \sum_{\ell_i, \ell_j \leq \ell_{\rm max}}
 \frac{\partial \ln C_W(\ell_i;\LAMBDA)}{\partial \ln \lambda_a}
\left[\bm{{\cal C}}^W\right]^{-1}_{\ell_i\ell_j}
 \frac{\partial \ln C_W(\ell_j;\LAMBDA)}{\partial \ln \lambda_b},
\label{fisher}
\EE
where $\lambda_a$ denote the $a$-th parameter;  $\lambda_1=A_s$ or
$\lambda_2=n_s$ in our definition. Note that we consider the window-convolved
power spectrum as the observable.
To calculate the power spectrum derivative, $\partial C_W(\ell)/\partial
 \ln\lambda_a$, we generated 100 realizations of the convergence maps, 
which are built based on the input linear power spectrum with $\pm 5\%$
 change of $\lambda_a$ on each side from its fiducial value
 (therefore 200 realizations in
 total). Then we evaluated the window-convolved power spectrum from the
average of the realizations, and used the spectra to
evaluate the derivatives $\partial C_W(\ell)/\partial \ln \lambda_a$
from the
 two-side numerical differentiation method. 
The fractional error on each parameter including marginalization over
 uncertainties of other parameter is given by
 $\Delta\lambda_a / \lambda_a = \sqrt{[\bm{F}]^{-1}_{aa}}$, where 
$[\bm{F}]^{-1}$ is the inverse of the Fisher matrix.

Fig.~\ref{fig_fisher} shows the errors of each parameter ($A_s$ or
$n_s$) expected for a hypothetical survey with 100 sq. degrees, but assuming
different survey geometries as in Fig.~\ref{fig:clw}.
As expected from the results of $S/N$ 
in Fig.~\ref{sn_rectang}, the most elongated geometry allows the
highest accuracy of these parameters over the range of $\ell_{\rm max}$
we consider. To be more precise, 
the elongated geometry of $0.5\times 200$ deg$^2$ gives about 3 or 25\%
improvement in the marginalized or unmarginalized error of $A_s$
at $\ell_{\rm max}\simeq 2000$, respectively,
compared to the square geometry of $10\times 10$ deg$^2$.
For $n_s$ the elongated geometry gives almost the same marginalized
error (more exactly speaking, 
0.3\% degraded error) and about 20\% improvement for the
unmarginalized error at $\ell_{\rm max}\simeq 2000$. 
Thus the improvement in the
error of $A_s$  is greater than that in the error of $n_s$.
However, the improvement in the marginalized error 
is milder compared to that in the unmarginalized error or 
the $S/N$ value, for $\ell_{\rm max}\simgt $a few $10^3$. 
Since the $S/N$ value is proportional to the volume of Fisher ellipse in
a multidimensional parameter space, the marginalized error is obtained
from the projection of the Fisher ellipse onto the parameter
axis, yielding a smaller improvement in the marginalized error \citep[see][for the similar discussion]{TakadaJain:09}.

\section{Sparse sampling optimization of the survey geometry}
\label{sec:sparse}

\begin{figure*}
\begin{center}
\includegraphics[width=150mm]{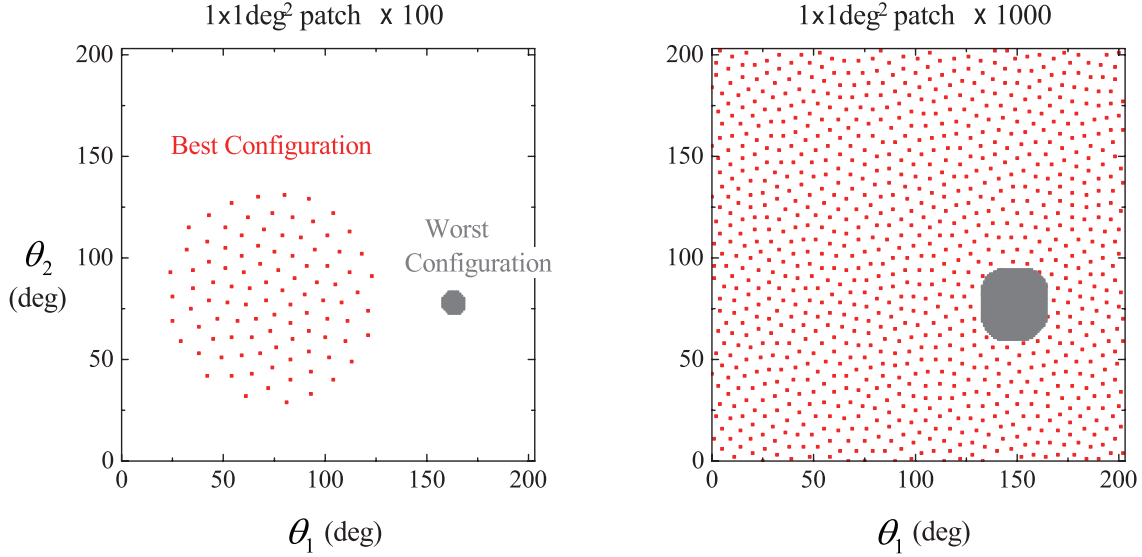}
\caption{The sparse-sampling strategy for the survey footprints when the
total area is fixed. The range shown is the all-sky map ($203\times 203$
sq. degrees) assuming the flat-sky approximation. 
Note that we assumed the periodic boundary condition beyond $203$ degrees.
Assuming that the
fundamental building block of the survey footprints is a square patch of
$1\times 1$ sq. degrees, we address which configuration of the $N_{\rm p}$
patches is best or worst in a sense that the configuration has the
smallest or largest SSC contamination, for the fixed total area of 100
(left panel) or 1000 (right) sq. degrees. 
For illustrative purpose, the best and worst configurations are plotted
within the same panel (in the right panel, the other patches are
similarly distributed under the worst configuration). Because of the
periodic boundary condition, the center position of each configuration
can be displaced in parallel.  For the best configuration, the different
patches are separated by about 15 degrees from each other. The angular
extent of all the patches is found to be about 10000 sq. degrees or all
sky ($203\times 203$ sq. degrees), respectively.  The worst
configuration is close to the square shape, with slightly rounded corners.
}
\label{fig_100&1000patches}
\end{center}
\end{figure*}
\begin{figure}
\begin{center}
\includegraphics[width=8cm]{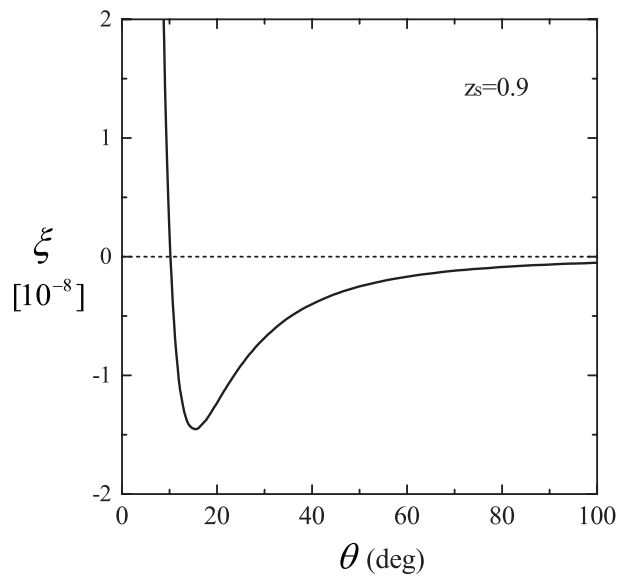}
\caption{The two-point correlation function of the log-normal
 convergence field, $\xi(\theta)$,  
for source redshift $z_s=0.9$. The two-point function has a negative
 minimum at about 15 degrees, which corresponds to the separation between
 the different patches for the best configuration in
 Fig.~\ref{fig_100&1000patches}.
}
\label{fig_xi}
\end{center}
\end{figure}
\begin{figure*}
\begin{center}
\includegraphics[width=160mm]{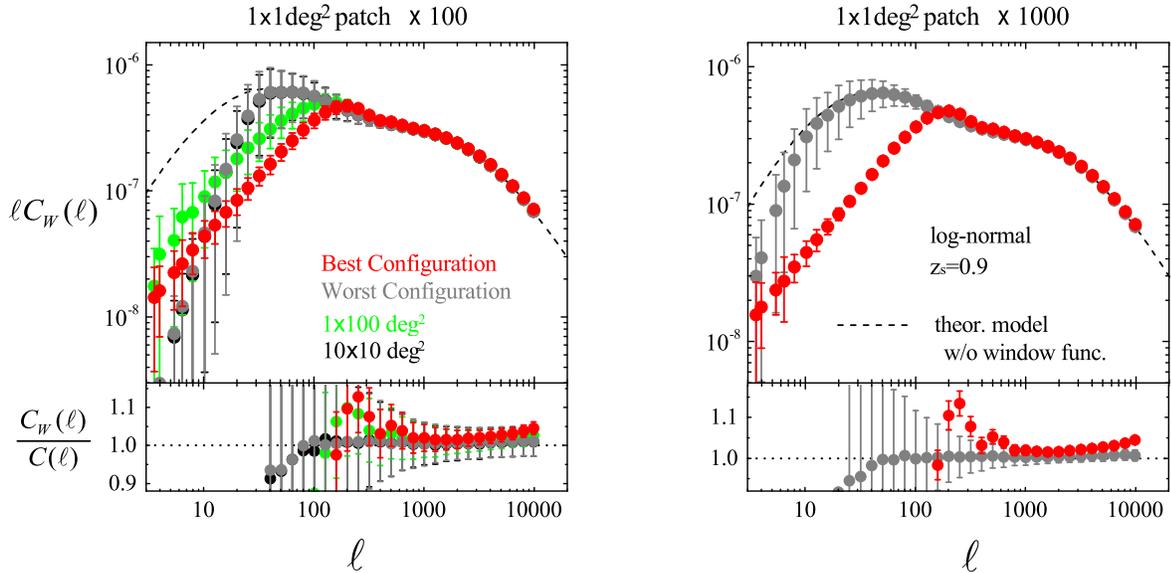}
\vspace*{-3cm}
\caption{The window-convolved power spectra for the best or worst
configuration for the fixed area of 100 or 1000 sq. degrees, for the
 sparse-sampling strategy as in Fig.~\ref{fig_100&1000patches}. 
In the left panel, for comparison, we also show the
result for the rectangular survey geometry of $1 \times 100$ sq. degrees
 (green) and the square geometry of $10 \times 10$ sq. degrees (black)
 in Fig.~\ref{fig:clw}. 
The error bar around each point denotes the $\pm 1\sigma$ scatters among the
 1000 realizations, clearly showing that the
scatter for the best configuration is smaller than that for the worst
configuration. The dashed curve is the true power spectrum. The lower
 plot shows the fractional difference compared to the true spectrum as
 in Fig.~\ref{fig:clw}.}  
\label{fig_cl_100&1000patches}
\end{center}
\end{figure*}
\begin{figure*}
\begin{center}
\includegraphics[width=150mm]{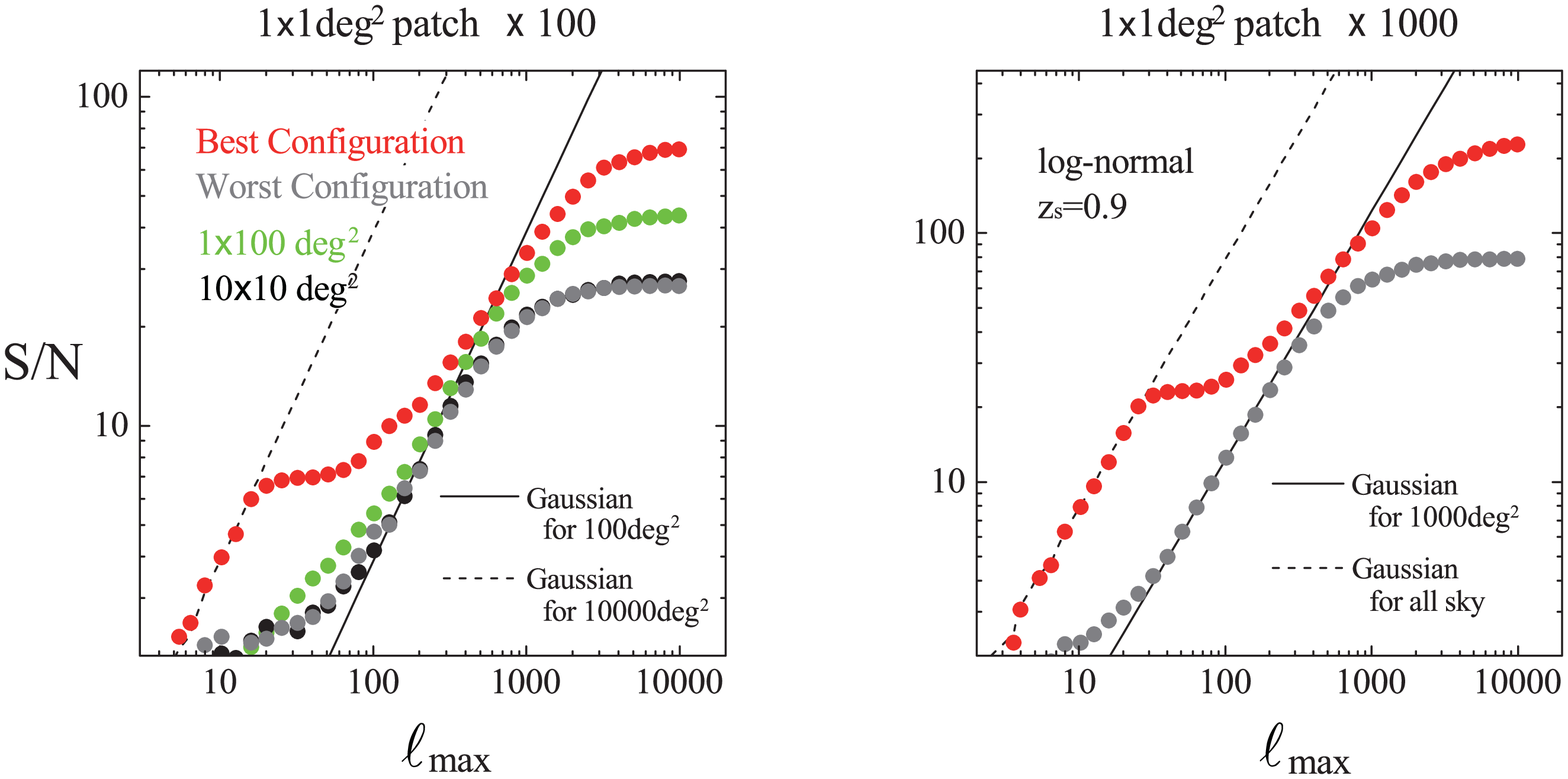}
\caption{The cumulative $S/N$ for the best or worst configurations in
Fig.~\ref{fig_100&1000patches}. In the left panel, for comparison, we
also show the results for the rectangular geometries of $1\times 100$
or $10\times 10$ sq. degrees. The dashed curve in each panel shows the
 $S/N$ value expected for a Gaussian field in the ``sparse-sampling''
regime, where we assumed the angular extent of the best configuration, 
$100$ or $203\times 203$ (all-sky) sq. degrees, as the ``effective'' survey
 area, respectively. To be more precise, we used the total number of
 Fourier modes around each multipole bin assuming the effective survey area:
$N_\ell = 2\upi\ell 
\Delta\ell/[(2\upi)^2/\Omega_{\rm eff}]$ and $\Omega_{\rm eff}$ is the
 effective area. On the other hand, 
the solid curve is the Gaussian prediction in the patch-inside regime,
 i.e. the value obtained assuming  the
 actual survey area, $100$ or 1000 sq. degrees. In the intermediate
 multipole range between the two regimes, the $S/N$ value appears to be
 constant, as no Fourier mode is sampled in this configuration. 
Nevertheless, the figure shows that the best configuration
 has the greater $S/N$ values over the wide range of multipoles. 
}
\label{fig_sn_100&1000patches}
\end{center}
\end{figure*}
We have so far considered a continuous geometry. 
In this section, we explore an optimal
sparse-sampling strategy. In this case the window function 
becomes even more complicated, causing a greater mixture between
different Fourier modes over a wider range. Observationally, a
continuous geometry might be to some extent preferred. There are
instances, where 
we want to avoid the mode coupling due to the window
function, especially in the presence of inhomogeneous selection function
over different pointings of a telescope. 
There are also instances, where we want to build a
continuous survey region by tiling different patches with an overlap
between different pointings, because such a strategy 
allows a better photometry calibration by comparing the
measured fluxes of the same objects in the overlapping regions 
across
the entire survey region 
\citep{Padmanabhanetal:08}. 
In addition the sparse sampling of the survey strategy
may require a more slewing of a telescope to
cover separated regions, which may cause an extra overhead and
therefore lower a survey efficiency for a given total amount of
the allocated observation time.
There are also instances, where we require 
a minimum size of a connected region
in order to have a sufficient sampling of the particular
Fourier mode such as the baryonic acoustic oscillation scale. Here we
ignore these possible observational disadvantages of a sparse sampling
strategy. Instead we here address a question:
what is the best sparse-sampling strategy for maximizing the information
 content of the power spectrum measurement for a fixed survey area?

Again recalling that the degradation in the power spectrum
measurement is mainly caused by the SSC effect, we can find 
the answer to the above
question by searching for a disconnected geometry that
minimizes $(\sigma_W)^2$ in Eq.~(\ref{eq:cov_unified}). For comparison,
we also search for the worst survey geometry in a sense that it gives
the lowest information content.
To find these geometries, we employ the following method.
First, we divide each map of the log-normal lensing field ($203 \times
203~ {\rm deg}^2$) into $203\times 203$ patches, i.e. each patch has an
area of 1 sq. degrees. Thus we consider each patch as the fundamental building
block of survey footprints for an assumed survey area\footnote{In the
following we use ``patch'' to denote the fundamental block of survey
footprints; here $1\times 1$ sq. degrees. On the other
hand, we use ``grid'' to denote the pixel of each simulated map, which
is $1\times 1$ sq. arcmin. Thus each patch contains $60 \times 60$ grids in our
 setting.}. 
The Subaru HSC
has a FoV of about 1.7 sq. degrees, so one may consider the FoV size
of a telescope for the patch. The following discussion can be applied
for any other size of the patch. 

In the following we assume either 100 or 1000 sq. degrees for the
total area, and then numerically search for configurations of 
the 100 or 1000 patches
which have the smallest or largest $(\sigma_W)^2$ value.
The numerical procedures are: 
\begin{itemize}
 \item[(i)]
Generate a random distribution of the $N_{\rm p}$ ($=100$ or $1000$) patches
 in the entire map ($203 \times 203$ patches in total).
\item[(ii)] Allow
the $i$-th patch's position to move to an unfilled patch, with fixing
other patches' positions, until 
the $i$-th patch's position yields the minimum or maximum value
 $(\sigma_W)^2$ computed from the total window function of $N_{\rm p}$ patches
based on Eq.~(\ref{sigma_w}).
\item[(iii)] Repeat the procedure (ii) for each of other patches iteratively
 (we may come back to the $i$-th patch)
 until the minimum or maximum $(\sigma_W)^2$ value is well converged.
 \item[(iv)] Redo the procedures (i)-(iii) from different initial
 distributions of the $N_{\rm p}$ patches.
\end{itemize}
We used $10^4$ initial positions.
In the following, we show the results for the best and worst configurations
obtained from the 10$^4$ initial positions, but we checked that the different
initial positions give almost the same configurations.

To make a fair comparison between different configurations/geometries,
we use the Fourier transform of the entire map region ($203\times 203$
deg$^2$); the patches outside the survey footprints or the unfilled
patches are zero-padded (i.e. set to $\kappa(\THE)=0$), and then perform
FFT with $12180^2$ grids to compute the Fourier-transformed field. In
this way, the fundamental Fourier mode (Fourier resolution) and the
maximum Fourier mode are the same for all the survey geometries.

Fig.\ref{fig_100&1000patches} shows the best and
worst configurations of the survey footprints
for each of 100 or 1000 sq. degrees, respectively.  
The best (worst) configuration has 
$(\sigma_W)^2=1.3 \times 10^{-7}$ ($8.8 \times 10^{-7}$) for $100~
{\rm deg}^2$ or $1.1 \times 10^{-8}$ ($1.0 \times 10^{-7}$)
for $1000~ {\rm deg}^2$, respectively.
For the best configuration, the distribution of the $N_{\rm p}$ patches (each
$1\times 1$ sq. degrees) appears regularly spaced, separated by $\sim
$15 deg. from each other, rather than random, as discussed below.
The angular extent of the best configuration is about 10000 sq. degrees
 or all-sky area (about 41000 sq. degrees) for the case of
100 or 1000 sq. degrees, respectively. Thus the filling fraction is only
1 or 2.4 per cent, respectively. Hence the sparse-sampling strategy
might allow for about factor 100 faster survey speed (equivalently
factor 100 less telescope time), 
compared to the 100 per cent filling strategy.  
On the other hand, the worst configuration is almost square
shaped, with slightly rounded corners.

To gain a more physical understanding of Fig.~\ref{fig_100&1000patches},
we can rewrite Eq.~(\ref{xi}) for $(\sigma_W)^2$ as
\BEA
 \left( \sigma_W \right)^2 = \frac{1}{\Omega_W^2} 
\sum_{i=1}^{N_p} \int \! d^2 \THE \!
 \int \! d^2 \THE^\prime \, W_{i}(\THE) W_{i}(\THE^\prime)
 \xi(|\THE-\THE^\prime|)  \nonumber \\
 + \frac{2}{\Omega_W^2} \sum_{i,j; i>j} \int \! d^2 \THE \!
 \int \! d^2 \THE^\prime \, W_{i}(\THE) W_{j}(\THE^\prime)
 \xi(|\THE-\THE^\prime|),
\EEA
where we re-defined the window function as $W(\THE)=\sum_i W_i(\THE)$,
and $W_i(\THE)$ is the window function of the $i$-th patch. Since
$W_i(\THE)=1$ when $\THE$ is inside the $i$-th patch, otherwise
$W_i(\THE)=0$, the first term arises from the integration of
$\xi(|\THE-\THE'|)$ when the vectors $\THE$ and $\THE'$ are in the same
patch. One the other hand, the second term arises form the integration
of $\xi(|\THE-\THE'|)$ when the vector $\THE$ and $\THE'$ are in the
different patches. As can be found from Fig.~\ref{fig_xi}, the first
term is always positive-additive, while the second term can have a
negative contribution, lowering $(\sigma_W)^2$, when the separation of
different patches is more than $\sim $10~deg. Since $\xi(r)$ has a
negative minimum at $r\sim 15$~deg., 
$(\sigma_W)^2$ can be minimized if 
taking a configuration so that 
different patches are separated by 
$\sim $15~deg. from each other.
Thus, even if different patches are separated
by an infinite angle, i.e. $\xi=0$, such a configuration does not have
the smaller $(\sigma_W)^2$. 

In Fig.~\ref{fig_cl_100&1000patches} we show the window-convolved power
spectra for the best and worst configurations. Compared to the true
power spectrum, the sparse sampling causes a significant change in the
convolved spectrum at $\ell\simlt $a few $10^2$, due to a significant
transfer of Fourier modes due to the complex window function. Here the
multipole scale of a few $10^2$ corresponds to the patch size ($1\times
1$ deg$^2$), the fundamental block of the survey footprints. At
multipoles 
$\ell\simgt $ a few $10^2$, the convolved power spectra become similar
to the true spectrum to within 5 per cent in the amplitude. 
As can be found from the lower panel, the best configuration clearly shows
the smaller scatter at each multipole bin among the 1000 realizations
than that of the worst configuration or more generally a
compact geometry.

Fig.~\ref{fig_sn_100&1000patches} shows the cumulative $S/N$ for the
best and worst configurations of 100 or 1000 sq. deg. area in
Fig.~\ref{fig_100&1000patches}.
The best configuration allows a higher $S/N$ of the power spectrum
measurement over the range of multipoles, from the linear to non-linear
regimes. 
Thus the sparse sampling allows an access to the larger angular (lower
multipole) scales \citep{kaiser98}.
For the case of 100 sq. degrees (the left
panel in Fig.~\ref{fig_100&1000patches}), the angular extent of the
different patches is about 10000 sq. degrees (or 100 deg. on a side).
The figure shows 
that the $S/N$ is close to the Gaussian expectation for the effective
area, $10000$ sq. degrees or all-sky area in the left or right panels,
respectively. To be more precise  the covariance matrix in this
regime is approximated as ${\cal
C}_{ij}=2C_W(\ell)^2\delta^{K}_{ij}/N_\ell$ with $N_\ell=2\upi \ell\Delta
\ell/[(2\upi)^2/\Omega_{\rm eff}]$, where $\Omega_{\rm eff}$ is the
effective area. 

The sparse-sampling,
by construction, can not probe Fourier modes over the range of 
intermediate angular scales such as $10\simlt \ell \simlt$ a few 
10$^2$ in our case. In this intermediate range, the $S/N$ value is flat and
does not increase with increasing $\ell_{\rm max}$. 
On the other hand, at the angular scales smaller than the
patch size ($\ell\simgt $a few $10^2$), the power spectrum measurement
arises from Fourier modes inside each patch. 
At the small scales, the SSC effect becomes significant. The figure
clearly shows that the best configuration allows for a factor 2 -- 2.5
greater $S/N$ at $\ell_{\rm max}\simgt 10^3$
than in the worst configuration. 
Also notice that, as can be found from the left panel,  
the best-configuration gives the higher $S/N$ than 
in the elongated rectangular geometry of $1\times 100$ sq. degrees,
whose shortest side length is the same as the patch size.  
Thus the sparse sampling strategy yields a higher precision of the power
spectrum measurement than a continuous geometry, for the fixed total area.

\begin{figure}
\begin{center}
\includegraphics[width=45mm]{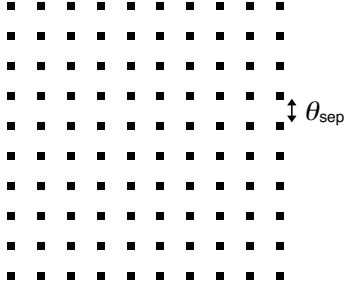}
\caption{Another working example of the sparse-sampling survey
footprints. Assuming that the fundamental building block is the patch
$1\times 1$ sq. degrees as in Fig.~\ref{fig_100&1000patches} and the
total survey area is 100 sq. degrees (100 patches), we study different
configurations of the 100 patches as a function of the separation angle 
$\theta_{\rm sep}$, as illustrated. 
}
\label{fig_10x10_patches}
\end{center}
\end{figure}
\begin{figure}
\begin{center}
\includegraphics[width=8cm]{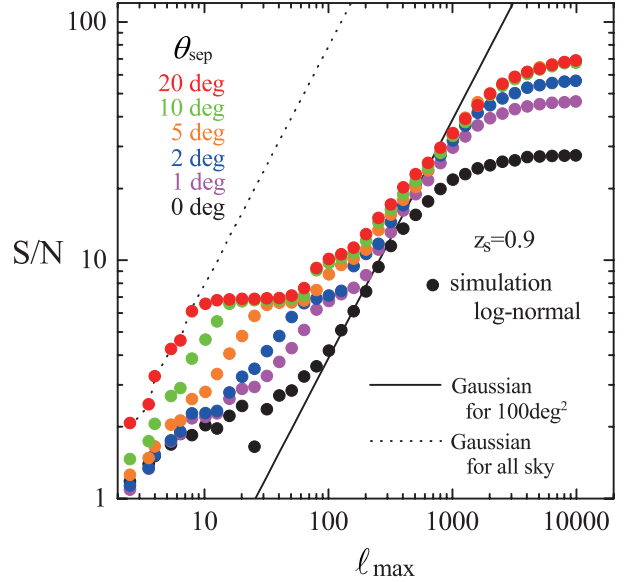}
\caption{The cumulative $S/N$ for the different configurations as a
function of the separation angle $\theta_{\rm sep}$ in
Fig.~\ref{fig_10x10_patches}. The dotted and solid curves show the $S/N$
 values expected for a Gaussian field, for the effective area of $100$
 or all sky, respectively. 
}
\label{fig_sn_thetasep}
\end{center}
\end{figure}
Finally, we further study the advantage of the sparse sampling strategy for the
power spectrum measurement. Assuming that the 100 patches (each patch is
$1\times 1$ sq. degrees) is regularly distributed and different patches
are regularly separated by the angle $\theta_{\rm sep}$ from each other
as given in Fig.~\ref{fig_10x10_patches}, Fig.~\ref{fig_sn_thetasep}  
shows how the $S/N$ value of power spectrum measurement changes with the
separation angle.
The continuous geometry, given by no separation ($\theta_{\rm
sep}=0$), yields the smallest $S/N$. 
The wider separation angle (larger $\theta_{\rm sep}$) allows an access to
the Fourier modes over the wider range of multipoles from the linear to
nonlinear regimes. If the separation angle is more than $5$
degrees, the SSC effect can be mitigated.

We have so far considered the fixed patch size, $1\times 1$ sq. degrees.
We have checked that, if the finer patch size is adopted for the fixed
total area, the best configuration further improves the total
information content of the power spectrum measurement over the wider
range of multipole bins. We also note that the results we have shown
qualitatively hold for different source redshifts, $z_s=0.6 $ -- 1.5.

\section{Conclusion and Discussion}
\label{sec:conclusion}

In this paper we have studied how the accuracy of weak lensing power
spectrum measurement varies with different survey geometries. We have
used the 1000 realizations of weak lensing maps and the analytical
model, assuming the log-normal model that approximates
non-Gaussian features seen in the weak lensing field for $\Lambda$CDM
model. Since the SSC effect arising from super-survey modes dominates
the non-Gaussian covariance in the range of $\ell\simeq 10^3$, the key
quantity to determine its survey geometry dependence is the variance of
the mean convergence mode in the survey region,
$(\sigma_W)^2=\ave{\bar{\kappa}_W^2}$, where
$\bar{\kappa}_W=(1/\Omega_S)\int\!d^2\THE W(\THE)\kappa(\THE)$.  We
showed that an optimal survey geometry can be found by looking for a
geometry to minimize $(\sigma_W)^2$ for a fixed total area. We used the
formulation in \cite{th13} to analytically derive the power spectrum
covariance and then used the analytical prediction to confirm the
finding from the simulated maps.

We showed that, for a fixed total area, the optimal survey geometry can
yield a factor 2 improvement in the cumulative $S/N$ of power spectrum
measurement, integrated up to $\ell_{\rm max}\simeq 10^3$, compared to
the $S/N$ in a compact geometry such as square and circular shaped
geometries. Furthermore, by taking a sparse sampling strategy, we can
increase the dynamic range of multipoles in the power spectrum
measurement, e.g., by a factor 100 in the effective survey area, if the
survey field is divided into 100 patches. Again, in this case, the
optimal survey design can be found by looking for a configuration of 100
patches to minimize the variance $(\sigma_W)^2$.

Our results might imply an interesting application for upcoming surveys.
For example, the LSST or Euclid surveys are aimed at performing
 an almost all-sky imaging
survey. If these surveys adopt a sparse-sampling strategy
with a few per cent filling factor  in the first
few years 
(Fig.~\ref{fig_100&1000patches}), 
the few per cent data might allow the
power spectrum measurements with an equivalent statistical precision
to that of the all-year data, i.e. enabling the desired cosmological
analysis very quickly. Then it can fill up unobserved
fields between the different patches in the following years. Thus,
while the same all-sky data is obtained in the end, taking a clever
survey strategy over years might allow for a quicker cosmological
analysis with the partial data in the early phase of the surveys.

In order to have the improved precision in the power spectrum
measurement with the optimal survey design, we need to properly
understand the effect of the survey window function. In reality,
inhomogeneous depth and masking effects need to be properly taken into
account. The sparse sampling
causes sidelobes in the Fourier-transformed window function, causing a
mixture of different Fourier modes in the power spectrum measurement
\citep[also see][]{kaiser98}. The effect of the side lobes also 
needs to be taken into account, when comparing the measurement with theory.
Throughout this paper we simply adopted the sharp window
function: $W(\THE)=0$ or 1 
 (see the sentences around Eq.~\ref{survey_area}).  To reduce the
mode-coupling due to the sharp window, we may want to use an apodization
of the window function, which is an operation to smooth out the sharp
window, e.g. with a Gaussian function, in order to filter out
high-frequency modes.
With such an apodization method, we can make 
the window-convolved power spectrum 
closer to the true power spectrum at a given multipole bin, which may be
desired in practice when comparing the measured power spectrum with theory.
However, the effective survey area decreases and it degrades the
extracted information content or the $S/N$ value at the price. Thus an
optimal window function needs to be explored depending on scientific
goals of a given survey.

Throughout this paper, we have employed the simple log-normal model to
approximate the weak lensing field in a $\Lambda$CDM model. 
We
believe that the results we have found are valid even if using the full
ray-tracing simulations. However, the brute-force approach requires
huge-volume $N$-body simulations to simulate a wide-area weak lensing
survey as well as requires the many realizations. This would be
computationally expensive. This problem can be studied by using a
hybrid method combining the numerical and analytical methods in
\citet{Lietal:14} and \citet{th13}. \citet{Lietal:14,Lietal:14b} showed that a
super-box mode can be included by introducing an apparent curvature
parameter $\Omega_{K}$, given in terms of the super-box mode $\delta_b$,
and then solving an evolution of $N$-body particles in the simulation
under the modified background expansion. As shown in \citet{th13}, since
the dependence of the SSC effect on survey geometry is determined mainly
by the variance $(\sigma_W)^2$, we can easily compute the variance by
using the analytical prediction for the input linear power spectrum
(Eq.~\ref{sigma_w}) or using the simulation realizations of linear
convergence field. Thus, by combining these methods, we can make a more
rigorous study of the survey geometry optimization for upcoming
wide-area surveys, at a reasonable computational expense.

Although we have studied the problem for a two-dimensional weak lensing
field, the method in this paper can be applied to a survey
optimization problem for a three-dimensional galaxy redshift survey.
Again various galaxy redshift surveys are being planned
\citep[e.g.][]{Takadaetal:14}, and the
projects are expensive both in time and cost, so the optimal survey
design is important to explore.

\section*{Acknowledgments}
We thank Chris Hirata, Wayne Hu, Atsushi Nishizawa and Ravi Sheth
for useful comments and discussions.
This work was supported in part by JSPS Grant-in-Aid for
Scientific Research (B) (No. 25287062) ``Probing the origin
of primordial mini-halos via gravitational lensing phenomena'', 
 and by Hirosaki University Grant for
Exploratory Research by Young Scientists.
 This work is also 
supported in part by Grant-in-Aid for Scientific Research from the JSPS
Promotion of Science (Nos. 23340061, 26610058, and 24740171), 
 and by World Premier
International Research Center Initiative (WPI Initiative), MEXT, Japan,
by the FIRST program ``Subaru Measurements of Images and Redshifts
(SuMIRe)'', CSTP, Japan.
MT  was also supported in part by the National Science Foundation
under Grant No. PHYS-1066293 and the
warm hospitality of the Aspen Center for
Physics.

\bibliographystyle{mn2e} \bibliography{mn-jour,refs} 

\onecolumn

\appendix

\section{Trispectrum of log-normal convergence field}
\label{app:4pt}

Let us consider the log-normal convergence field $\kappa$:
 its mean is zero and its statistics is characterized by the two-point
 correlation function, $\xi_{12} \equiv \xi(|\THE_1-\THE_2|) =
 \langle \kappa(\THE_1) \kappa(\THE_2) \rangle$.
Then, the four-point correlation of $\kappa$ can be written in terms
 of $\xi$ as (Hilbert et al. 2011),
\begin{eqnarray} 
 && \langle \kappa(\THE_1) \kappa(\THE_2) \kappa(\THE_3) \kappa(\THE_4)
 \rangle =  \xi_{12} \xi_{34} + \xi_{13} \xi_{24} + \xi_{14} \xi_{23} 
 + \frac{1}{\kappa_0^2} \left[ \xi_{12} \xi_{13} \xi_{14}
 + \xi_{12} \xi_{13} \xi_{24} + \xi_{12} \xi_{13} \xi_{34}
 + \xi_{12} \xi_{14} \xi_{23} + \xi_{12} \xi_{14} \xi_{34}
 \right. \nonumber \\
 &&~  + \xi_{12} \xi_{23} \xi_{24}
 + \xi_{12} \xi_{23} \xi_{34} + \xi_{13} \xi_{14} \xi_{23} 
 + \xi_{13} \xi_{14} \xi_{24} + \xi_{13} \xi_{23} \xi_{24}
 + \xi_{13} \xi_{23} \xi_{34} + \xi_{14} \xi_{23} \xi_{24}
 + \xi_{14} \xi_{23} \xi_{34} + \xi_{12} \xi_{24} \xi_{34} \nonumber \\
 &&~ \left. + \xi_{13} \xi_{24} \xi_{34} + \xi_{14} \xi_{24} \xi_{34}
 \right] + O \left( \xi^4/\kappa_0^4 \right)
\end{eqnarray}
The first three terms are disconnected parts, while the others are connected
 parts which are leading correction terms arising from non-Gaussianity. 
We ignore the higher-order terms since $\xi/\kappa_0^2$ is usually very small.
This approximation corresponds to ``the simplified log-normal approximation''
 in Hilbert et al. (2011). 
By performing Fourier transform, we have  
\begin{eqnarray}
 && \langle \tilde{\kappa}(\ELL_1) \tilde{\kappa}(\ELL_2)
 \tilde{\kappa}(\ELL_3) \tilde{\kappa}(\ELL_4) \rangle =
 C(\ell_1) C(\ell_3) \delta_{\rm D}^2(\ELL_{12})
 \delta_{\rm D}^2(\ELL_{34}) + C(\ell_1) C(\ell_2) \delta_{\rm D}^2(\ELL_{13})
 \delta_{\rm D}^2(\ELL_{24}) + C(\ell_1) C(\ell_2) \delta_{\rm D}^2(\ELL_{14})
 \delta_{\rm D}^2(\ELL_{23}) \nonumber \\
 &&~ + \frac{\left( 2\upi \right)^2}{\kappa_0^2}
 \left[ C(\ell_1) C(\ell_2) C(\ell_3) +
 C(\ell_1) C(\ell_2) C(\ell_4) + C(\ell_1) C(\ell_3) C(\ell_4) +
 C(\ell_2) C(\ell_3) C(\ell_4) + C(\ell_2) C(\ell_4) C(\ell_{12})
  \right. \nonumber \\
 &&~  + C(\ell_2) C(\ell_3) C(\ell_{12}) + C(\ell_3) C(\ell_4) C(\ell_{13})
 + C(\ell_2) C(\ell_3) C(\ell_{13}) + C(\ell_3) C(\ell_4) C(\ell_{14})
 + C(\ell_2) C(\ell_4) C(\ell_{14})  \nonumber \\
 &&~ + C(\ell_1) C(\ell_2) C(\ell_{23}) + C(\ell_1) C(\ell_3) C(\ell_{23})
 + C(\ell_1) C(\ell_2) C(\ell_{24}) + C(\ell_1) C(\ell_4) C(\ell_{24})
 + C(\ell_1) C(\ell_4) C(\ell_{34})
  \nonumber \\  
 &&~ \left. + C(\ell_1) C(\ell_3) C(\ell_{34})
  \right] \delta_{\rm D}^2(\ELL_{1234}) + O \left( C^4/\kappa_0^4 \right),
\end{eqnarray}
where $\ELL_{ij} = \ELL_i + \ELL_j$ and $\ELL_{ijkl}=\ELL_i + \ELL_j +
 \ELL_k + \ELL_l$.
The trispectrum is defined as the connected part of the above function,
 $\langle \tilde{\kappa}(\ELL_1) \tilde{\kappa}(\ELL_2) \tilde{\kappa}(\ELL_3)
 \tilde{\kappa}(\ELL_4) \rangle_{\rm c} = (2 \upi)^2
 T(\ELL_1,\ELL_2,\ELL_3,\ELL_4) \delta_{\rm D}^2(\ELL_{1234})$.
Then we have,
\begin{eqnarray}
 && T(\ELL_1,\ELL_2,\ELL_3,\ELL_4) = \frac{1}{\kappa_0^2}
 \left[ C(\ell_1) C(\ell_2) C(\ell_3) +
 C(\ell_1) C(\ell_2) C(\ell_4) + C(\ell_1) C(\ell_3) C(\ell_4) +
 C(\ell_2) C(\ell_3) C(\ell_4)   \right. \nonumber \\
 &&~  + C(\ell_2) C(\ell_4) C(\ell_{12}) + C(\ell_2) C(\ell_3) C(\ell_{12})
 + C(\ell_3) C(\ell_4) C(\ell_{13})
 + C(\ell_2) C(\ell_3) C(\ell_{13}) + C(\ell_3) C(\ell_4) C(\ell_{14})
   \nonumber \\
 &&~ + C(\ell_2) C(\ell_4) C(\ell_{14}) + C(\ell_1) C(\ell_2) C(\ell_{23})
 + C(\ell_1) C(\ell_3) C(\ell_{23})
 + C(\ell_1) C(\ell_2) C(\ell_{24}) + C(\ell_1) C(\ell_4) C(\ell_{24})
  \nonumber \\  
 &&~ \left. + C(\ell_1) C(\ell_4) C(\ell_{34})
   + C(\ell_1) C(\ell_3) C(\ell_{34}) \right].
\end{eqnarray}
In a particular configuration of $\ELL_1+\ELL_2=\ELL_3+\ELL_4=0$, 
 the trispectrum has a simple form,
\begin{equation}
 T(\ELL_1,-\ELL_1,\ELL_2,-\ELL_2) = \frac{1}{\kappa_0^2}
 \left[ 2 C(\ell_1) C(\ell_2) \left\{ C(\ell_1) + C(\ell_2) \right\} 
 + \left\{ C(\ell_1) + C(\ell_2) \right\}^2
 \left\{ C(|\ELL_1+\ELL_2|) +  C(|\ELL_1-\ELL_2|) \right\} \right]. 
\label{trisp}
\end{equation}

\twocolumn

\section{Results: $\Lambda$CDM ray-tracing simulations}
\label{app:sato}

\begin{figure}
\begin{center}
\includegraphics[width=8cm]{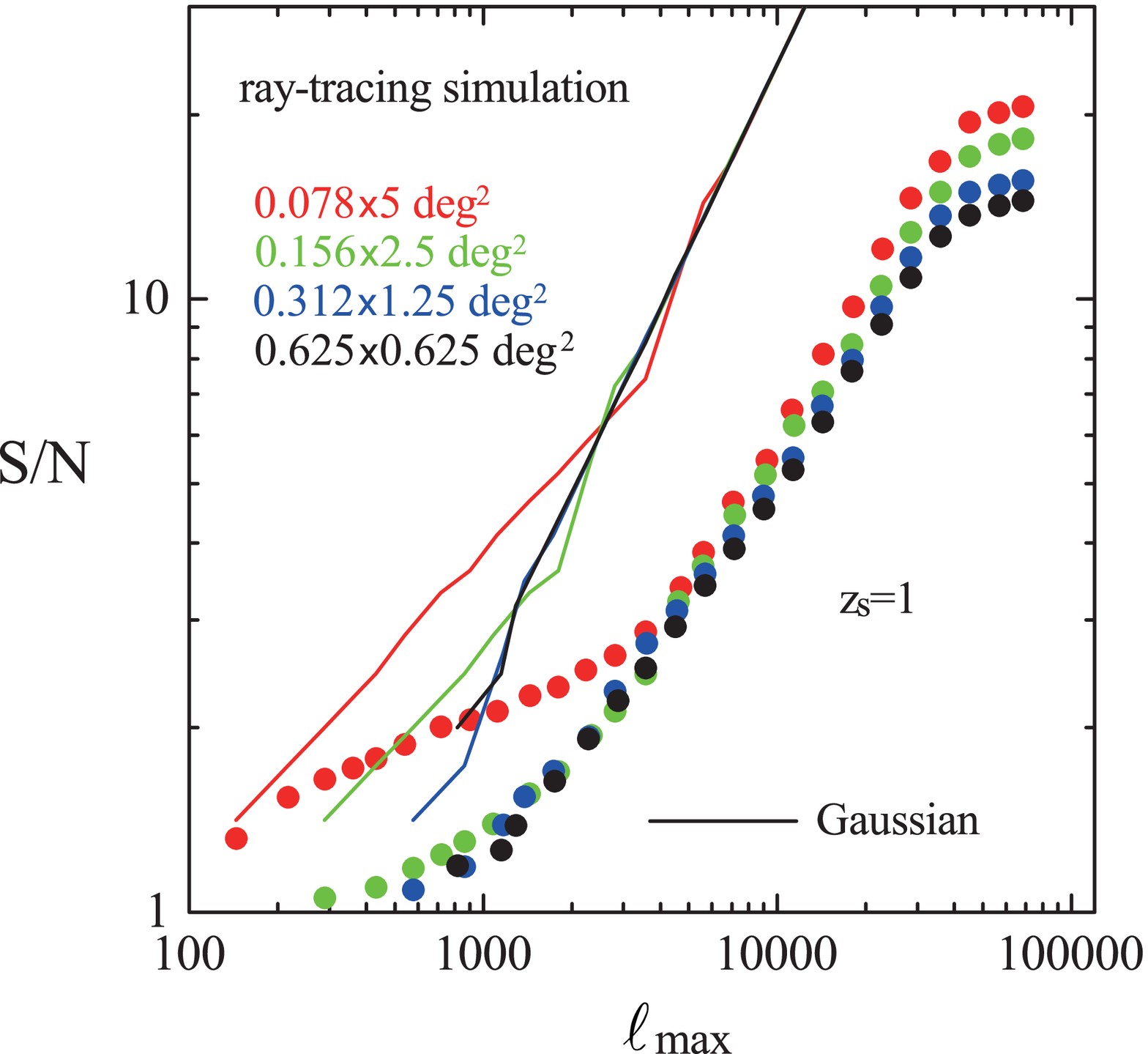}
\caption{
The cumulative $S/N$ measured from the ray-tracing simulations for
 $\Lambda$CDM model and source redshift $z_s=1$, taken from
 \citet{sato09}. 
The different color symbols show the results for different geometries:
the square-shaped geometry with area  $0.625 \times 0.625~ {\rm deg}^2$
 (black), and the rectangular-shaped geometries with the same area, but
 with different side ratios, 
$0.078 \times 5$ (red),
 $0.156 \times 2.5$ (green), and
 $0.312 \times 1.25$ deg$^2$ (blue), respectively. 
The solid curves are the Gaussian error predictions.
}
\label{fig_sato_sn}
\end{center}
\end{figure}

In most part of this paper we have used the simulated convergence maps
for the log-normal model. Our method allows us to simulate the
convergence field over a wide area (all-sky area), thereby including all
the Fourier modes from very small scales to all-sky scales, and to
simulate many realizations at a computationally cheap cost. However, the
log-normal model is an empirical model to mimic the lensing field for a
$\Lambda$CDM model.
In this
appendix we use the ray-tracing simulations in \citet{sato09} to study
whether the results we show hold for a more realistic lensing field.

Each of the 1000 realizations in \citet{sato09} has an area of $5\times
5$ sq. degrees in square shaped geometry, and is given in $2048^2$ grids
(each grid size is 0.15 arcmin on a side). As can be found from Fig.~1
in \citet{sato09}, the ray-tracing simulations were done in a light cone
of area $5 \times 5 {\rm deg}^2$, viewed from an observer position (z = 0). 
The projected mass density fields in intermediate-redshift slices were
generated from N-body simulations which have a larger simulation box
than the volume covered by the light cone. Hence the lensing fields have
contributions from the mass density field of scales outside the
ray-tracing simulation area, although, exactly speaking, the modes
outside the N-body simulation box were not included. Thus the
ray-tracing simulations include the SSC effect. As discussed in
Section~3 of \citet{sato09}, the ray-tracing simulation would not be so
reliable at $\ell\simgt 6000$ due to the resolution issue of the
original N-body simulations. However, since we are interested in the
effect of different survey geometries, we below use the simulations down
to the pixel scale.

Although the ray-tracing simulation map is in small area ($5\times 5$
deg$^2$), we want to study a wide range of different geometries
available from the simulated map. Here we consider a square-shaped
geometry of $0.625\times 0.625(=0.39)$ sq. degrees and
rectangular-shaped geometries of different side ratios: $0.078 \times
5$, $0.156 \times 2.5$, and $0.312\times 1.25$ deg$^2$, which have the
side ratios of 1:64, 1:16 and 1:4, respectively. Thus these areas are
much smaller than that of planned weak lensing surveys. For this small
area, the SSC effect arises from the average convergence mode in the
nonlinear regime, rather than the linear regime, and the SSC
contribution relative to the standard covariance terms is relatively
smaller than expected for a wider area survey \citep[see Fig.~1
in][]{th13}. Thus the dynamic range of different
geometries is smaller than in the log-normal simulations, where we
studied down to 1:400 ratio.  Fig.~\ref{fig_sato_sn} shows the
cumulative $S/N$ for the different geometries. For this plot, we used
the 1000 realizations for source redshift $z_s=1$.
The covariance matrix is reliably estimated by using the 1000 realizations. 
The multipole range we studied is all in the nonlinear
regime, due to the small area ($0.39$ sq. degrees). For comparison, the
solid curves show the $S/N$ values expected for the Gaussian field for
each geometry, which is computed by accounting for the number of Fourier
modes available for each multipole bin. All the simulation results are
much below the Gaussian expectation, meaning that the non-Gaussian
errors significantly degrade the $S/N$ value over the range of
multipoles.  Comparing the results for different geometries shows a
clear trend that the more elongated geometry yields a higher $S/N$
value; about 40\% higher $S/N$ value at $\ell_{\rm max}\simeq 2000$ in
the $0.078\times 5$ deg$^2$  than in the $0.625\times 0.625$ deg$^2.$
Thus these results qualitatively confirm our finding based on the log-normal
distribution. To check these for a wider area comparable with that of
upcoming surveys requires ray-tracing simulations done for a much wider area.

\end{document}